\documentclass[11pt, a4paper]{article}
\pdfoutput=1
\usepackage{jheparxiv}
\usepackage[latin1]{inputenc}
\usepackage{amsmath}
\usepackage{amsfonts}
\usepackage{amssymb}
\usepackage{latexsym}
\usepackage{mathrsfs}
\usepackage{graphicx}
\usepackage{color}
\usepackage{twistor}
\usepackage[all]{xy}

\renewcommand{\d}{\mathrm{d}}

\subheader{\hfill \texttt{DAMTP-2013-63}}

\title{Worldsheet factorization for twistor-strings}

\author{Tim Adamo}

\affiliation{Department of Applied Mathematics \& Theoretical Physics \\
        University of Cambridge \\
        Wilberforce Road \\
        Cambridge CB3 0WA, United Kingdom}
	
\emailAdd{t.adamo@damtp.cam.ac.uk}

\abstract{We study the multiparticle factorization properties of two worldsheet theories which--at tree-level--describe the scattering of massless particles in four dimensions: the Berkovits-Witten twistor-string for $\cN=4$ super-Yang-Mills coupled to $\cN=4$ conformal supergravity, and the Skinner twistor-string for $\cN=8$ supergravity.  By considering these string-like theories, we can study factorization at the level of the worldsheet before any Wick contractions or integrals have been performed; this is much simpler than considering the factorization properties of the amplitudes themselves.  In Skinner's twistor-string this entails the addition of worldsheet gravity as well as a formalism that represents all external states in a manifestly symmetric way, which we develop explicitly at genus zero.  We confirm that the scattering amplitudes of Skinner's theory, as well as the gauge theory amplitudes for the planar sector of the Berkovits-Witten theory, factorize appropriately at genus zero.  In the non-planar sector, we find behavior indicative of conformal gravity in the Berkovits-Witten twistor-string.  We contrast factorization in twistor-strings with the story in ordinary string theory, and also make some remarks on higher genus factorization and disconnected prescriptions.}

\begin{document}

\maketitle


\section{Introduction}

Scattering amplitudes are highly constrained by their behaviour under multiparticle factorization, which dictates their singularity structure in the infrared.  Understanding these factorization properties has significantly improved our ability to constrain and compute scattering amplitudes in perturbative quantum field theory over the past twenty years.  At tree-level, this is perhaps most evident in the BCFW recursion relations \cite{Britto:2004ap, Britto:2005fq}, which allow us to construct the entire classical S-matrix of Yang-Mills theory or gravity by simply applying Cauchy's theorem to the amplitudes and using the analytic properties enforced by multiparticle factorization.  At loop level, the power of factorization is manifest in the modern unitarity approach \cite{Bern:1994zx, Bern:1994cg}, which constructs loop amplitudes from on-shell tree amplitudes by looking at `cuts' where internal loop propagators go on-shell (see \cite{Bern:2011qt, Britto:2010xq, Carrasco:2011hw} for reviews).   

In string theory, scattering amplitudes are also constrained by their factorization properties.  While this can be demonstrated by studying factorization limits of amplitude formulae (c.f., \cite{D'Hoker:1988ta}), the algebro-geometric nature of worldsheet gravity allows us to study factorization at the level of the worldsheet before any Wick contractions or integrals have occurred.  In particular, the $p^{-2}$ poles corresponding to an internal propagator going on-shell are manifested in string theory by simple poles in the moduli variables near the boundary of the moduli space.  At this boundary divisor, the worldsheet is pinched and there is a sum over states propagating across the cut; it is precisely on-shell physical states which lead to the appropriate simple pole in the moduli.  This \emph{worldsheet factorization} was first described by Polchinski for the bosonic string \cite{Polchinski:1988jq}, and has been emphasized more recently in Witten's formulation of superstring perturbation theory \cite{Witten:2012bh}.  In essence, any string theory which is anomaly free and has reasonable vertex operators will automatically have the correct factorization properties by virtue of the geometry in play.

Ordinarily, we regard conventional scattering amplitudes in gauge theory or gravity as emerging from the low-energy ($\alpha'\rightarrow 0$) limit of string theory amplitudes.  However, it has been realized that these gauge and gravity amplitudes can themselves be written in a fashion which is highly reminiscent of a string-theory origin--at least at tree-level.  These developments stemmed from Witten's twistor-string theory (and its first-order formulation due to Berkovits) \cite{Witten:2003nn, Berkovits:2004hg}, which describes the S-matrix of planar $\cN=4$ super-Yang-Mills (SYM) theory in terms of a string theory whose target space is twistor space.  At genus zero, this led to an explicit formula of Roiban, Spradlin, and Volovich (RSV) for the entire tree-level S-matrix of planar $\cN=4$ SYM \cite{Roiban:2004yf} whose validity was confirmed by studying its factorization properties \cite{Skinner:2010cz, Dolan:2011za} (i.e., the formula was shown to satisfy BCFW recursion).  The gravitational vertex operators of the Berkovits-Witten twistor-string have been shown to correspond to non-minimal $\cN=4$ conformal supergravity \cite{Berkovits:2004jj, Ahn:2005es, Dolan:2008gc, Adamo:2012nn}, so beyond genus zero the gauge theory amplitudes of the theory are contaminated by conformal gravity degrees of freedom running in the loops.  

More recently, a formula for the tree-level S-matrix of $\cN=8$ supergravity in terms of maps from rational curves to twistor space was discovered by Cachazo and Skinner \cite{Cachazo:2012kg}, and its veracity determined by again demonstrating multiparticle factorization \cite{Cachazo:2012pz}.  Subsequently, Skinner found a worldsheet theory whose genus zero scattering amplitudes correspond to this formula \cite{Skinner:2013xp}, raising the possibility that higher loop supergravity amplitudes could be computed using this novel twistor-string at higher genus.  

Given the RSV or Cachazo-Skinner formula, one might ask: what is the utility of having a worldsheet theory which produces the formula at genus zero?  In this paper, we provide one answer to this question: multiparticle factorization for these formulae is obtained organically using the worldsheet theory.  This is significantly easier than proving factorization at the level of the formula itself (one need only glance at the factorization calculations in \cite{Skinner:2010cz, Dolan:2011za, Cachazo:2012pz} to confirm this).  Understanding this worldsheet factorization also provides insight into potentially novel structures at tree-level (such as the MHV formalism \cite{Cachazo:2004kj}), and raises the possibility of extending the formulae to higher genus.   

In ordinary string theory, these factorization properties follow immediately from worldsheet gravity and the geometry of the moduli space.  Hence, it may seem that our motivations for studying worldsheet factorization in the Berkovits-Witten and Skinner models are rather trivial: surely factorization follows because these are string theories.  However, the story for these theories is not so simple.  Skinner's theory does not actually contain worldsheet gravity, and in both theories the vertex operators are different from the states appearing in ordinary string theory.  In particular, \emph{all} vertex operators in twistor-string theory are specified by generic cohomology classes on the worldsheet, so all `physical' states automatically have conformal weight zero.  Additionally, the Berkovits-Witten model contains non-simple poles in the factorization limit when considering multi-trace contributions to the scattering amplitudes away from the planar limit.  Such higher-order poles correspond precisely to the expected conformal gravity modes of the theory \cite{Berkovits:2004jj}.

This should be contrasted against the story in ordinary string theory, where a vertex operator having conformal weight zero is translated into a statement about its associated momentum being on-shell.  Furthermore, a higher-order pole in the factorization limit of string theory is related to the presence of tachyonic states rather than higher-order poles in momentum space. 

Furthermore, factorization requires setting up a calculational framework in which all external states are represented by the same type of vertex operator (i.e., fixed vertex operators).  We refer to this as a `manifestly permutation symmetric' formalism, and in Skinner's theory it requires the introduction of Picture Changing Operators.  Hence, some subtlety is required to study worldsheet factorization in twistor-string theory, even at genus zero!

\medskip

We begin in Section \ref{Theories} by introducing the two theories we study.  This includes setting out each worldsheet theory and describing its vertex operators in detail.  Here, the main issue is to set up a formalism in which scattering amplitudes are computed in a totally permutation symmetric way at the level of the worldsheet; or in other words, all external states are represented by the same type of vertex operator.  This should be contrasted with the procedure for studying factorization at the level of the amplitude, where the appropriate permutation invariance is built into the answer.  We describe this formalism in general, and compute the genus zero scattering amplitudes explicitly.

For gauge theory amplitudes in the Berkovits-Witten twistor-string, worldsheet permutation invariance is immediate since all external states are represented by vertex operators of the form $\tr(a\wedge j)$, where $a$ is the $\cN=4$ SYM multiplet in twistor space and $j$ is a gauge current on the worldsheet.  The genus zero scattering amplitudes with Yang-Mills external states then take the familiar RSV form in the planar sector \cite{Roiban:2004yf}:
\be{RSVi}
\cA_{n,d}=\int\frac{\prod_{a=0}^{d}\d^{4|4}U_{a}}{\mathrm{vol}\;\GL(2,\C)} \tr\left(\prod_{i=1}^{n}\frac{a_i\;\D\sigma_{i}}{(\sigma_{i}\sigma_{i+1})}\right),
\ee
where the integral is over the moduli space of holomorphic maps of degree $d$ from the Riemann sphere with $n$ marked points to twistor space $\PT\subset\P^{3|4}$.  The parameters $\{U_{a}\}$ are the map moduli, and $\{\sigma_i\}$ are homogeneous coordinates on $\P^1$ with $\SL(2,\C)$-invariant inner product $(\sigma_{i}\sigma_{j})\equiv\epsilon_{\alpha\beta}\sigma_{i}^{\alpha}\sigma_{j}^{\beta}$.  The weight two holomorphic differential on $\P^1$ is denoted $\D\sigma=(\sigma\;\d\sigma)$.  The quotient by the volume of $\GL(2,\C)$ is meant in the usual Fadeev-Popov sense, and accounts for the $\SL(2,\C)$ automorphism group of the Riemann sphere and the $\C^{*}$ rescalings of the map moduli.

In Skinner's twistor-string, amplitudes are also computed by integrating over the moduli space of degree $d$ holomorphic maps to twistor space, but now from a worldsheet with fermionic structure \cite{Skinner:2013xp}.  Anomaly cancellation for the theory fixes a twistor space with $\cN=8$ supersymmetry, and worldsheet correlators must account for the new fermionic automorphisms and deformations which were absent in the Berkovits-Witten theory.  Genus zero amplitudes in Skinner's twistor-string are given by the Cachazo-Skinner formula \cite{Cachazo:2012kg}:
\be{CSi}
\cM_{n,d}=\int\frac{\prod_{a=0}^{d}\d^{4|8}U_{a}}{\mathrm{vol}\;\GL(2,\C)}\mathrm{det}'\left(\HH\right)\;\mathrm{det}'\left(\HH^{\vee}\right)\prod_{i=1}^{n}h_i\;\D\sigma_i,
\ee
where $h_i$ are the $\cN=8$ supergravity external states on twistor space, the matrices $\HH$, $\HH^{\vee}$ are the Hodges and dual Hodges matrices respectively, and the reduced determinants $\mathrm{det}'$ are given by
\begin{equation*}
\mathrm{det}'\left(\HH\right)=\frac{\left|\HH^{1\cdots d+2}_{1\cdots d+2}\right|}{|\sigma_{1}\cdots\sigma_{d+2}|^2}, \qquad \mathrm{det}'\left(\HH^{\vee}\right)=\frac{\left|\HH^{\vee\;d+1\cdots n}_{\;d+1\cdots n}\right|}{|\sigma_{1}\cdots\sigma_{d}|^{2}}.
\end{equation*}
In this notation, $\HH^{i\cdots j}_{i\cdots j}$ stands for the matrix $\HH$ with the rows and columns $i,\ldots,j$ removed, and the denominator factors are the Vandermonde determinants:
\begin{equation*}
|\sigma_{1}\cdots\sigma_{d+2}|=\prod_{i<j\in\{1,\ldots, d+2\}}(\sigma_{i}\sigma_{j}).
\end{equation*}

While it has been shown that the Cachazo-Skinner formula is permutation invariant with respect to the external states, the correlator which leads directly to \eqref{CSi} is not manifestly permutation symmetric at the level of the worldsheet.  In particular, $d+2$ of the external states are represented by \emph{fixed} vertex operators, while the remaining $n-d-2$ take the form of \emph{integrated} vertex operators.  In a manifestly permutation symmetric setup, all $n$ external states should be represented by fixed vertex operators.  This entails introducing appropriate picture changing operators to ensure that there is still a top-degree form on the fermionic moduli space.  We show how to do this in Section \ref{SkTS}; actually computing the genus zero scattering amplitudes then leads to the alternative formula:
\be{CS2i}
\cM_{n,d}=\int\frac{\prod_{a=0}^{d}\d^{4|8}U_{a}}{\mathrm{vol}\;\GL(2,\C)}\frac{\left|\mathsf{H}\right|}{\left|\mathsf{N}\right|^2}\frac{\left|\HH^{\vee}\right|}{|y_{1}\cdots y_{d}|^{2}}\prod_{i=1}^{n}h_i,
\ee
which is \emph{a priori} dependent on a set of $n-d-2$ points $x_{j}\in\P^1$ and $d$ points $y_{k}\in\P^1$ that stand for the locations of picture changing operators. The matrix $\mathsf{H}$ is a $(n-d-2)\times(n-d-2)$ generalization of the Hodges matrix, while $|\mathsf{N}|$ is a Slater determinant depending on both $\sigma_i$ and $x_j$.  $\HH^{\vee}$ is a $d\times d$ dual Hodges matrix depending only on the $y_{k}$ positions.  

Despite the apparent dissimilarities, we show that \eqref{CS2i} is actually \emph{equal} to the Cachazo-Skinner formula and is furthermore independent of the $x_j$ and $y_k$.  This demonstrates explicitly that the permutation invariant formalism produces the correct answer at the level of the scattering amplitudes, while at the same time providing us with a setup appropriate for studying factorization at the level of the worldsheet.  

In Section \ref{WSG}, we describe worldsheet gravity and the worldsheet factorization limit for these theories.  The anomaly-free Berkovits-Witten twistor-string contains chiral worldsheet gravity, while Skinner's twistor-string contains no worldsheet gravity at all, although it does have fields corresponding to a worldsheet gravitino.  Hence, we insert a $bc$-ghost system into the theory by hand, at the price of a conformal anomaly.  We argue that this does not affect the study of worldsheet factorization at genus zero, but will cause problems at higher genus.  In describing the worldsheet factorization limit, we take our cue from Witten's exposition in section 6 of \cite{Witten:2012bh}.

Section \ref{GZero} studies worldsheet factorization for genus zero scattering amplitudes in these theories.  In Skinner's theory and the planar limit of the Berkovits-Witten theory, we find factorization behaviour consistent with unitarity.  This provides an alternative proof of the RSV and Cachazo-Skinner formulae.  In particular, we see that factorization emerges immediately as a consequence of our permutation symmetric setup and worldsheet gravity, so this proof is substantially simpler than proving factorization at the level of \eqref{RSVi} or \eqref{CSi}.  

In the non-planar sector of the Berkovits-Witten twistor-string, we find a factorization channel that has a double pole corresponding to conformal supergravity.  Away from the planar limit, there are multi-trace terms which contribute to the amplitudes of the twistor-string.  Double poles arise from factorization channels that do not disturb the structure of the worldsheet current algebra correlator, which is already cut by the multiple traces.  Clearly, this behavior is different from what usually happens in string theory, so we include some discussion contrasting twistor-strings with ordinary string theory at the level of worldsheet factorization in Section \ref{StrComp}. 

Finally, Section \ref{Concl} concludes with a discussion of our findings, open questions, and future directions.  In particular, we give some heuristic remarks about factorization at higher genus in these theories, and consider the implications of genus zero factorization for deriving disconnected formalisms (e.g., a MHV vertex expansion for gravity).  We also make some comments on the potential for applying worldsheet factorization to the study of new formulae for the scattering of gluons and gravitons in arbitrary dimension (c.f., \cite{Cachazo:2013hca}), as well as leading singularities for $\cN=8$ supergravity.  

Three appendices contain technical details associated with the arguments in this paper.  Appendix \ref{GZFs} contains a derivation of the manifestly permutation symmetric S-matrix of Skinner's theory at genus zero, and a proof of its equivalence to the Cachazo-Skinner formula.  Appendix \ref{QProps} reviews the role of the modulus controlling the factorization limit in twistor-string theory.  We also study the factorization properties of multi-trace contributions to the Berkovits-Witten theory at the level of the amplitude in Appendix \ref{DTF}.


\section{Worldsheet Theories and Permutation Symmetric Setup}
\label{Theories}

In this section, we introduce the two worldsheet theories we will study for the remainder of this paper.  For each, we describe the worldsheet theory at the level of the action and its BRST symmetries, and then give the vertex operators.  In Skinner's twistor-string it is most natural to calculate scattering amplitudes in a formalism where some vertex operators are \emph{fixed} and the remainder are \emph{integrated} \cite{Skinner:2013xp}.  This is due to the fermionic structure of the worldsheet, which is a $(1|2)$-dimensional split supermanifold.  While this mixture of fixed and integrated vertex operators is convenient for calculating explicit worldsheet correlators, it is not a good formalism to use for studying worldsheet factorization since the external states are not represented symmetrically.  This issue is familiar from traditional (super)string theory, where integrated vertex operators are useful for practical computations but can lead to problems at non-generic momenta (c.f., \cite{Witten:2012bh}).

To consider worldsheet factorization, we need to put all external states on the same footing at the level of the worldsheet correlator--before any Wick contractions or integrals have been performed.  This entails using fixed vertex operators for all external states, at the cost of introducing \emph{Picture Changing Operators} (PCOs) which ensure that we still obtain a top-degree form on the fermionic moduli space.  Generally speaking, PCOs are BRST-closed operators which are nearly BRST-exact; they are inserted in the worldsheet path integral to fix fermionic moduli, a process which can be interpreted as making a choice for the vacuum of the superconformal $\beta\gamma$-ghost system.  The role of PCOs can also be understood from the perspective of super-geometry (e.g., \cite{Belopolsky:1997jz}).  They were first introduced in superstring perturbation theory by \cite{Friedan:1985ey, Friedan:1985ge} and further developed in \cite{Witten:1986qs, Verlinde:1986kw}.  In this paper, we will use the representation of a PCO due to Verlinde \& Verlinde \cite{Verlinde:1987sd}, where for a general $\beta\gamma$-system the operator is given by:
\be{genPCO}  
\Upsilon=\{Q,\Theta(\beta)\}=\delta(\beta)\;\{Q,\beta\},
\ee
with $Q$ the relevant BRST operator and $\Theta$ the Heavyside step function.

Constructing these operators for the Skinner theory allows us to set up scattering amplitudes which are manifestly permutation symmetric with respect to external states at the level of the worldsheet.  To illustrate how the formalism works, we give the genus zero scattering amplitudes in this permutation invariant setup, obtaining a generalized representation of the Cachazo-Skinner.  This formula is treated in detail by Appendix \ref{GZFs}.

The reader who is already familiar with these theories may wish to simply skim this section, moving on the the discussion of worldsheet gravity and factorization that follows.


\subsection{Berkovits-Witten theory}

Witten first formulated twistor-string theory as a topological B model on $\cN=4$ twistor space (an open subset $\PT\subset\P^{3|4}$) supplemented with $D1$-instantons \cite{Witten:2003nn}.  Berkovits' subsequently reformulated this theory in terms of a first-order worldsheet action for an open string with boundary on the real slice $\RP^{3|4}\subset\P^{3|4}$, and the role of the $D1$-instantons replaced by the more conventional worldsheet instantons \cite{Berkovits:2004hg}.  These two theories are equivalent at the level of their perturbation theory, so we refer to them jointly as the `Berkovits-Witten twistor-string.'  In practice, we will use a closed string analogue of Berkovits' model, which is based on the heterotic twistor-string developed in \cite{Mason:2007zv}; this avoids the necessity of explicit choices of reality structure on twistor space and makes the formalism somewhat cleaner.  A detailed treatment of this theory was recently given by Reid-Edwards \cite{ReidEdwards:2012tq}, and we gloss over many of those details here.  However, the essential structures needed for studying worldsheet factorization are the same, since the twistor-string has chiral (holomorphic) worldsheet gravity. 

Let $\Sigma$ be a closed (Euclidean) Riemann surface, endowed with a line bundle $\cL\rightarrow\Sigma$ of degree $d$.  The matter fields of the Berkovits-Witten twistor-string are
\be{BWmatt}
Z\in\Omega^{0}(\Sigma,\C^{4|4}\otimes\cL), \qquad Y\in\Omega^{0}(\Sigma,\C^{4|4}\otimes K_{\Sigma}\otimes\cL), \qquad \alpha\in\Omega^{0,1}(\Sigma),
\ee
where $K_{\Sigma}$ denotes the canonical bundle (we will denote the tangent bundle by $T_{\Sigma}=K^{-1}_{\Sigma}$).  The field $Z$ has the interpretation of a degree $d$ map $Z^{I}:\Sigma\rightarrow\C^{4|4}$ with dual variables $Y_{I}$, while $\alpha$ is a $\GL(1,\C)$ gauge field on the worldsheet associated with the line bundle $\cL$.  The matter action is then given by
\be{BWma}
S[Z,Y,\alpha]=\frac{1}{2\pi}\int_{\Sigma}Y_{I}\;\dbar Z^{I}+\alpha\wedge Y_{I}\;Z^{I},
\ee
to which we also add the action for a worldsheet current algebra, denoted $S_{C}$, for the Lie algebra $\mathfrak{g}$ of a gauge group.

This action is invariant under a local $\C^{*}$ symmetry acting on the matter fields:
\begin{equation*}
Z^{I}\rightarrow \e^{\gamma}Z^{I}, \qquad Y_{I}\rightarrow\e^{-\gamma}Y_{I}, \qquad \alpha\rightarrow\alpha-\dbar\gamma,
\end{equation*}   
so gauging this symmetry reduces the target space from $\C^{4|4}$ to an open subset of the Calabi-Yau supermanifold $\P^{3|4}$.  To gauge-fix this action, we introduce the usual Virasoro ghost $c\in\Pi\Omega^{0}(\Sigma, T_{\Sigma})$ and anti-ghost $b\in\Pi\Omega^{0}(\Sigma,K^{2}_{\Sigma})$, as well as a $\GL(1,\C)$ ghost system $v\in\Pi\Omega^{0}(\Sigma)$, $u\in\Pi\Omega^{0}(\Sigma,K_{\Sigma})$.\footnote{Throughout this paper, $\Pi$ denotes parity-reversing functor: for some bundle $\mathcal{S}$, $\Pi\Omega^{0}(\mathcal{S})$ denotes the space of sections of $\mathcal{S}$ which are fermionic-valued.}  This leaves us with the full worldsheet action for the Berkovits-Witten theory:
\be{BWaction}
S=\frac{1}{2\pi}\int_{\Sigma}Y_{I}\;\dbar Z^{I}+b\;\dbar c+u\;\dbar v+S_{C}.
\ee

The relevant anomalies for this theory are the $\GL(1,\C)$ anomaly $\mathfrak{a}_{\GL(1)}$ (due to the fields which couple to the line bundle $\cL$) and the central charge $\mathfrak{c}$.  The former vanishes because the only $\GL(1,\C)$-charged fields are $Y_{I}$ and $Z^{I}$, which have an equal number of bosonic and fermionic components (projectively), since the target space is a subset of $\P^{3|4}$.  Likewise, the contribution from the $YZ$-system to the central charge vanishes, leaving only
\begin{equation*}
\mathfrak{c}=-26-2+\mathfrak{c}_{C},
\end{equation*}
where $\mathfrak{c}_{C}$ is the central charge of the current algebra.  Hence, the theory is anomaly free when $\mathfrak{c}_{C}=+28$.  Furthermore, its BRST operator is given by \cite{Berkovits:2004tx, Dolan:2007vv}:
\be{BWbrst}
Q=\oint v\;Y_{I}\;Z^{I}+c\;Y_{I}\;\partial Z^{I}-c\;b\;\partial c+\frac{3}{2}\partial^{2}c-c\;u\;\partial v+c\;T_{C},
\ee
where $T_{C}$ is the stress-energy tensor of the worldsheet current algebra.  This can be shown to obey $Q^2=0$.  

As in any string theory, vertex operators for this theory are given by marginal deformations of the action which are BRST closed.  The first such vertex operator couples to the worldsheet current algebra, and corresponds to the gauge-theoretic degrees of freedom in the theory:
\be{BWVa}
V^{a}=\int_{\Sigma}\tr\left(a(Z)\wedge j\right),
\ee
where $a\in\Omega^{0,1}(\Sigma,\mathfrak{g})$ and $j\in\Omega^{0}(\Sigma, K_{\Sigma}\otimes\mathfrak{g})$.  This obeys $\{Q,V^{a}\}=0$ since $a$ is homogeneous with respect to $Z^{I}$, and furthermore encodes the full $\cN=4$ SYM multiplet for the Lie algebra $\mathfrak{g}$ via the Penrose transform of $a\in H^{0,1}(\PT,\cO\otimes\mathfrak{g})$.  

Since this theory has a Calabi-Yau target, its gravitational vertex operators correspond to deformations of the complex and Hermitian structures of the target space.  These are represented respectively by $f\in\Omega^{0,1}(\PT, T_{\PT})$ and $g\in\Omega^{0,1}(\PT,T^{*}_{\PT})$, and are translated into marginal deformations of the action by the vertex operators:
\be{BWVfg}
V^{f}=\int_{\Sigma}Y_{I}\;f^{I}(Z), \qquad V^{g}=\int_{\Sigma} g_{I}(Z)\;\partial Z^{I},
\ee
where $f^{I}\in\Omega^{0,1}(\Sigma, \cL)$ is constrained to be volume preserving ($\partial_{I}f^{I}=0$) and $g_{I}\in\Omega^{0,1}(\Sigma,\cL^{-1})$ obeys $Z^{I}g_{I}=0$ since $f$ is defined up to multiples of the Euler vector field on $\PT$.  These properties can be used to confirm that $\{Q,V^{f}\}=\{Q,V^{g}\}=0$.  $\cV^{f}$ and $\cV^{g}$ encode the field content of $\cN=4$ conformal supergravity, via either half-Fourier transform \cite{Berkovits:2004jj} or the Penrose transform \cite{Adamo:2013cra}.

For this theory, all vertex operators are of the same type (i.e., an integral over the worldsheet $\Sigma$); this means that correlation functions are manifestly permutation symmetric at the level of the worldsheet with respect to external states.  In this paper, we will consider scattering amplitudes of the Berkovits-Witten twistor-string where all external states are represented by the gauge theory vertex operators $V^{a}$ given in \eqref{BWVa}; the single-trace contribution (i.e., planar limit of $\mathfrak{g}$) contribution to the genus zero amplitude is the RSV formula:
\be{RSV}
\cA_{n,d}=\int\frac{\prod_{a=0}^{d}\d^{4|4}U_{a}}{\mathrm{vol}\;\GL(2,\C)}\left\la \prod_{i=1}^{n}V_{i}^{a}\right\ra =\int\frac{\prod_{a=0}^{d}\d^{4|4}U_{a}}{\mathrm{vol}\;\GL(2,\C)} \tr\left(\prod_{i=1}^{n}\frac{a_i\;\D\sigma_{i}}{(\sigma_{i}\sigma_{i+1})}\right).
\ee
Here, the zero-modes of the map $Z^{I}:\Sigma\rightarrow\PT$ are given by 
\begin{equation*}
Z^{I}(\sigma)=\sum_{a=0}^{d}U_{a}^{I}(\sigma^{0})^{a}(\sigma^{1})^{d-a},
\end{equation*}
and the quotient by $\mathrm{vol}\;\GL(2,\C)$ accounts for the $\SL(2,\C)$ automorphism group of $\Sigma\cong\P^{1}$ and the $\C^{*}$ rescalings of the homogeneous coordinates $\sigma$ and map coefficients $U_{a}$.  So the integrals remaining in \eqref{RSV} correspond to integration over the moduli space $\overline{M}_{0,n}(\PT,d)$ of holomorphic maps of degree $d$ from $\P^{1}$ with $n$ marked points to $\PT$.\footnote{More formally, this space can be understood as a supersymmetric analogue of Kontsevich's moduli space of stable maps.  At the level of the moduli stack, it is well-defined and shares the desirable properties of its bosonic cousin (i.e., the super-geometric version of the Deligne-Mumford property) \cite{Adamo:2012cd}.}


\subsection{Skinner theory}
\label{SkTS}

The fundamental difference between Skinner's twistor-string and the Berkovits-Witten theory is the presence of super-geometric structures at the level of the worldsheet \cite{Skinner:2013xp}.  In particular, the worldsheet of Skinner's theory is a $(1|2)$-dimensional split supermanifold $\rX$ defined by a closed Riemann surface $\Sigma$ and a rank-2 sheaf of superalgebras given by 
\be{SA}
\cD\cong\Pi\left(\C^{2}\otimes K^{-\frac{1}{2}}_{\Sigma}\otimes\cL\right),
\ee
where $\cL\rightarrow\Sigma$ is once more a line bundle of degree $d\geq 0$.  Note that since $\{\cD,\cD\}\subset\cD$, it is clear the $\rX$ is not a $\cN=2$ super-Riemann surface; the split condition means that
\begin{equation*}
T \rX=T\Sigma\oplus\cD,
\end{equation*}
so we can work with coordinates $(\sigma,\theta^{a})$ for a local fermionic coordinate $\theta^{a}$ and $a=1,2$.

As a supermanifold, $\rX$ has some rather attractive geometric properties.  Its holomorphic Berezinian sheaf obeys $\Ber(\rX)\cong\cL^{\otimes 2}$, and when the genus of the underlying bosonic worldsheet $\Sigma$ is zero it is given by a weighted projective superspace: $\rX_{g=0}\cong\mathbb{WP}^{1|2}_{(1,1|d+1,d+1)}$.  Its super-geometry also means that $\rX$ has non-trivial automorphism and deformation moduli even when $\Sigma$ does not.  For instance, fermionic automorphisms of $\rX$ are generated by sections $V\in\Gamma(\cD)$, which can be expanded as
\be{autos}
V=\left(v^{a}+\theta^{b}R^{a}_{b}+\frac{\theta^2}{2}\tilde{v}^{a}\right)\frac{\partial}{\partial\theta^{a}},
\ee
where
\begin{equation*}
v\in\Gamma(K^{-\frac{1}{2}}_{\Sigma}\otimes\cL), \qquad R\in\Gamma(\mathrm{End}(\C^{2}\otimes\cL)), \qquad \tilde{v}\in\Gamma(K^{\frac{1}{2}}_{\Sigma}\otimes\cL^{-1}).
\end{equation*}
Since the line bundle $\cL$ has degree $d\geq0$, we see that $\tilde{v}$ cannot be a globally holomorphic section (generically), so the only automorphisms that need to be accounted for are those encoded by $v$ and $R$.  A short Riemann-Roch calculation reveals that the number of global sections of $v$ is $(d+2-2g)$, which is the dimension of the space of corresponding fermionic automorphism zero-modes that will need to be integrated over in a worldsheet correlation function. 

Additionally, we must account for deformations of $\rX$ which act only on the superstructure of the worldsheet.  As a complex supermanifold, infinitesimal deformations of $\rX$ are given by elements of
\begin{equation*}
\mathrm{Def}(\rX)\cong\Ext^{1}_{\rX}(\Ber(\rX),\cO_{\rX})\cong H^{1}(\rX,T_{\rX}),
\end{equation*}
by Grothendieck duality.  So the deformations which parametrize only the odd moduli of $\rX$ are given by $H^{1}(\rX,\cD)\subset H^{1}(\rX,T_{\rX})$.  Serre duality for supermanifolds \cite{Penkov:1983, Ogievetskii:1984, Haske:1987} tells us that
\begin{equation*}
H^{1}(\rX,\cD)\cong H^{0}(\rX,\Ber(\rX)\otimes\cD^{\vee})^{\vee}\cong\Pi H^{0}(\rX,K^{\frac{1}{2}}_{\Sigma}\otimes\cL)^{\vee}.
\end{equation*}
Of course, as long as $d>0$ this space will have dimension $d$, so there are non-trivial fermionic moduli for the worldsheet itself.

The matter fields of Skinner's theory are encoded in a single superfield $\mathcal{Z}\in\Omega^{0}(\rX,\C^{4|8}\otimes\cL)$, which can be expanded as
\begin{equation*}
\mathcal{Z}^{I}=Z^{I}(\sigma)+\theta^{a}\rho_{a}^{I}(\sigma)+\frac{\theta^2}{2}Y_{I}(\sigma),
\end{equation*}
where $Z$ and $Y$ have the same interpretation as in the Berkovits-Witten twistor-string, and $\rho_{a}\in\Pi\Omega^{0}(\Sigma,\C^{4|8}\otimes K^{\frac{1}{2}}_{\Sigma})$ are a pair of worldsheet spinors.  The matter action is then given by specifying a non-degenerate skew-symmetric form $\la \cdot,\cdot\ra$, called the `infinity twistor', and taking
\begin{equation*}
S[\mathcal{Z}]=\frac{1}{4\pi}\int_{\rX}\la\mathcal{Z},\bar{\D}\mathcal{Z}\ra.
\end{equation*}
Here $\bar{\D}=\dbar+\Gamma$ is a worldsheet covariant derivative with $\Gamma\in\Omega^{0,1}(\rX,\cD)$.  This is gauge-fixed by introducing a ghost superfield $C\in\Pi\Omega^{0}(\rX,\cD)$ and its anti-ghost partner $B\in\Pi\Omega^{0}(\rX,\Ber(\rX)\otimes\cD^{\vee})$ with action
\begin{equation*}
S[B,C]=\frac{1}{2\pi}\int_{\rX}B_{a}\;\dbar C^{a}.
\end{equation*}
As in the Berkovits-Witten theory, gauging an overall $\C^{*}$-rescaling reduces the target space from $\C^{4|8}$ to $\PT\subset\P^{3|8}$.

In this paper, we will be concerned only with the flat-space version of Skinner's twistor-string.  This entails taking the limit where the infinity twistor becomes degenerate, and is accomplished by implementing matter field redefinitions using the dual infinity twistor, $[\cdot,\cdot]$ 
\begin{equation*}
Z^{I}\rightarrow Z^{I}, \qquad \rho^{I}_{1}\rightarrow \rho^{I}, \qquad \rho^{I}_{2}\rightarrow [\cdot,\tilde{\rho}], \qquad Y^{I}\rightarrow [\cdot, Y],
\end{equation*}
and then rescaling matter action by an inverse power of the cosmological constant \cite{Skinner:2013xp}.  Written in terms of the component fields, this leaves us with the gauge-fixed matter action we will consider in this paper:
\be{Smatt}
S[Z,\rho,\tilde{\rho},Y]=\frac{1}{2\pi}\int_{\Sigma}Y_{I}\;\dbar Z^{I}+\tilde{\rho}_{I}\;\dbar\rho^{I}.
\ee
Likewise, we can expand the ghost superfields $B,C$ in components as:
\begin{eqnarray*}
B_{a} & = & \mu_{a}+\theta^{b}\left(\epsilon_{ab}\mathrm{m}+\mathrm{m}_{ab}\right)+\frac{\theta^2}{2}\beta_{a}, \\
C^{a} & = & \gamma^{a}+\theta^{b}\left(\delta^{a}_{b}\frac{\mathrm{n}}{2}+\mathrm{n}^{a}_{b}\right)+\frac{\theta^2}{2}\nu^{a},
\end{eqnarray*}
leaving us with the ghost action:
\be{Sgh}
S[B,C]=\frac{1}{2\pi}\int_{\Sigma}\beta_{a}\dbar\gamma^{a}+\mathrm{m}_{ab}\dbar \mathrm{n}^{ab}+\mathrm{m}\;\dbar\mathrm{n}+\mu_{a}\dbar\nu^{a}.
\ee
The full worldsheet theory is described by the sum of \eqref{Smatt} and \eqref{Sgh}.

In addition to the potential $\GL(1,\C)$ and central charge anomalies, this theory also has potential $\SL(2,\C)$ and mixed $\GL(1,\C)$/ gravitational anomalies.  With respect to $\mathfrak{a}_{\GL(1)}$, the matter fields $Y$, $Z$ now contribute $-4$ due to the $\cN=8$ supersymmetry of $\PT$; the other fields which are charged under $\cL$ are $\beta\gamma$ and $\mu\nu$-systems, each of which contribute $+2$ (since they are bosonic), giving $\mathfrak{a}_{\GL(1)}=0$.  Similar calculations show that $\mathfrak{a}_{\SL(2)}=\mathfrak{a}_{\GL(1)/\mathrm{grav}}=0$.  For the central charge, one uses the standard dictionary to compute:
\begin{equation*}
\mathfrak{c}=2\times (-4) _{YZ}-4_{\tilde{\rho}\rho}+2\times 11_{\beta\gamma}-2\times 4_{\mathrm{m}\mathrm{n}}+2\times (-1)_{\mu\nu}=0.
\end{equation*}

This theory has a BRST operator given by:
\begin{multline}\label{Sbrst}
Q=\oint \gamma^{1}Y_{I}\rho^{I}+\gamma^{2}[Y,\tilde{\rho}]+\frac{\nu^{1}}{2}\la\rho, Z\ra+\frac{\nu^{2}}{2}\tilde{\rho}_{I}Z^{I} +\frac{\mathrm{n}}{2}Y_{I}Z^{I} \\
+\frac{1}{2}(\mathrm{n}^{12}+\mathrm{n}^{21})\rho^{I}\tilde{\rho}_{I}+\frac{\mathrm{n}^{11}}{2}\la\rho,\rho\ra+\frac{\mathrm{n}^{22}}{2}[\tilde{\rho},\tilde{\rho}]+\beta_{a}\left(\frac{\mathrm{n}}{2}\gamma^{a}+\mathrm{n}^{a}_{b}\gamma^{b}\right) \\
+\mu_{a}\left(\mathrm{n}^{a}_{b}\nu^{b}-\frac{\mathrm{n}}{2}\nu^{a}\right)+\mathrm{m}\gamma^{a}\nu_{a}-\mathrm{m}_{ab}\left(\mathrm{n}^{(a}_{c}\mathrm{n}^{b)c}+\gamma^{(a}\nu^{b)}\right),
\end{multline}
where we use the notation $\la\cdot ,\cdot\ra=\epsilon^{\alpha\beta}$, and $[\cdot,\cdot]=\epsilon_{\dot{\alpha}\dot{\beta}}$.  One can show that $Q^{2}=0$ is also satisfied, so \eqref{Sbrst} gives a well-defined BRST cohomology for building vertex operators.

\medskip

In contrast to the Berkovits-Witten twistor-string, vertex operators in Skinner's theory must take into account the fermionic automorphisms and deformations of the worldsheet supermanifold $\rX$.  To begin, consider the fermionic automorphisms appearing in \eqref{autos}; global sections generating these automorphisms correspond to zero modes of the ghost field $\gamma_{a}\in\Omega^{0}(\Sigma, K^{-\frac{1}{2}}_{\Sigma}\otimes\cL)$, of which there are $d+2-2g$.  In order to fix these automorphisms we must require that the translations generated by them act trivially at $d+2-2g$ points in the bosonic worldsheet $\Sigma$.

At the level of the path integral over $\gamma$, we can fix each point $\sigma_{i}\in\Sigma$ by inserting
\begin{equation*}
\delta^{2}(\gamma_{i})\equiv\delta(\gamma^{1}(\sigma_i))\;\delta(\gamma^{2}(\sigma_i)).
\end{equation*}
Since $\gamma_{a}$ takes values in $K^{-\frac{1}{2}}_{\Sigma}\otimes\cL$, it follows that we can interpret $\delta^{2}(\gamma)$ as an element of $\Omega^{0}(\Sigma, K_{\Sigma}\otimes\cL^{\otimes-2})$.  Hence, to build a well-defined operator on $\Sigma$ we must pair the delta-functions in $\gamma$ with some $h\in\Omega^{0,1}(\Sigma, \cL^{\otimes 2})$.  This leads us to the vertex operators:
\be{VO}
V^{h}=\int_{\Sigma}\delta^{2}(\gamma)\;h(Z).
\ee
Note that since $h$ takes values in $\cL^{\otimes 2}$ on the worldsheet, it follows that $h\in\Omega^{0,1}(\PT,\cO(2))$.  This means that (on-shell), $h$ encodes the $\cN=8$ gravity multiplet via the Penrose transform.  

However, only $d+2-2g$ insertions of $V^{h}$ are required to fixed all the fermionic automorphisms.  In Skinner's original formulation, the remaining $n-d-2+2g$ external states in a $n$-point scattering amplitude are then represented by \emph{integrated} vertex operators \cite{Skinner:2013xp}
\be{intVO}
\widehat{V}^{h}=\int_{\rX}h(\mathcal{Z})=\int_{\Sigma}\left[Y,\frac{\partial h(Z)}{\partial Z}\right]+\left[\tilde{\rho}, \frac{\partial}{\partial Z}\left(\rho^{I}\frac{\partial h(Z)}{\partial Z^{I}}\right) \right].
\ee
Performing calculations with this mixture of vertex operators at genus zero leads to the Cachazo-Skinner formula for the tree-level S-matrix of $\cN=8$ supergravity \cite{Cachazo:2012kg} in a relatively straightforward fashion.  In particular, this `mixed' formalism seems to be optimal from a practicality standpoint: the answer is \emph{the} optimal formula for the scattering amplitudes.   

This prescription should actually be familiar from traditional superstring theory, where a mixture of fixed and integrated vertex operators is optimal for explicitly computing scattering amplitudes \cite{D'Hoker:1988ta, Witten:2012bh}.  A similar story exists for the pure spinor formalism \cite{Berkovits:2000fe}, although slightly more care is required in constructing the measure due to some subtleties in the path integral over the pure spinor due to poles in ghost insertions (c.f., \cite{Grassi:2009fe, Aisaka:2009yp, Berkovits:2013pla}).      

While the use of integrated vertex operators simplifies the actual computation of scattering amplitudes, it poses a problem for studying worldsheet factorization.  In particular, all external states in the worldsheet correlator are not on the same footing \emph{a priori}.  The $d+2-2g$ vertex operators \eqref{VO} contain an insertion of $\delta^{2}(\gamma)$, which fixes the fermionic location of the external state on $\rX$, while the remaining operators of the form \eqref{intVO} are integrated over all of $\rX$.  To study worldsheet factorization, we want to represent all $n$ external states by the vertex operators $V^h$. 

So we want to compute a $n$-point amplitude by inserting $n$ of the operators \eqref{VO} for external states.  This enlarges the space of fermionic automorphisms to have dimension $n$, so we need a mechanism for reducing this space to its correct dimension of $d+2-2g$.  This is accomplished by inserting $n-d-2+2g$ \emph{Picture Changing Operators} (PCOs) for the $\beta\gamma$-ghost system.  In Skinner's theory $\beta_{a}$ take values in $K^{\frac{3}{2}}_{\Sigma}\otimes\cL^{-1}$, so using \eqref{genPCO} the relevant PCO is:
\be{bPCO*}
\widetilde{\Upsilon}=2\delta^{2}(\beta)\;\{Q,\beta^{a}\}\;\{Q,\beta_{a}\},
\ee
where the BRST operator \eqref{Sbrst} gives
\begin{equation*}
\{Q,\beta_{a}\}=\la\rho^{a},Y\ra-\frac{\mathrm{n}}{2}\beta_{a}-\mathrm{n}^{b}_{a}\beta_{b}-\mathrm{m}\nu_{a}-\mathrm{m}_{ab}\nu^{b},
\end{equation*}
before taking the flat-space limit.  Since there is an overall factor of $\delta^{2}(\beta)$, we can drop any terms proportional to $\beta$ from the current in the PCO leaving us with the following expression for the flat-space PCO:
\be{bPCO}
\widetilde{\Upsilon}=\delta^{2}(\beta)\left( Y_{I}\rho^{I}\;[Y,\tilde{\rho}]+\cdots\right),
\ee
where the dots represent terms which do not have zero modes at genus zero and can therefore be dropped for the purposes of this paper.

Additionally, we must include PCOs for the $\mu\nu$-system which account for the fermionic deformations corresponding to the $2d$ zero modes of ghost $\mu_{a}$.  Using the prescription of \eqref{genPCO} gives:
\be{mPCO*}
\Upsilon= 2\delta^{2}(\mu)\;\{Q,\mu^{a}\}\;\{Q,\mu_{a}\},
\ee
with
\begin{equation*}
\{Q,\mu_{a}\}=\frac{1}{2}\la Z,\rho_{a}\ra+\frac{\mathrm{n}}{2}\mu_{a}-\mathrm{n}_{a}^{b}\mu_{b}+\mathrm{m}\gamma_{a}+\mathrm{m}_{ab}\gamma^{b}.
\end{equation*}
Taking the flat-space limit leaves us with
\be{mPCO}  
\Upsilon=\delta^{2}(\mu)\left(\la\rho,Z\ra\;\tilde{\rho}_{I}Z^{I}+\cdots\right),
\ee
where the dots again represent terms which have no zero modes at genus zero. Clearly, these PCOs are required whether or not we choose to work with integrated vertex operators.  Indeed, they play a crucial role in calculating the scattering amplitudes in that picture by constructing the `dual Hodges matrix' factor of the Cachazo-Skinner formula \cite{Skinner:2013xp}.

In summary, we have set up a formalism for Skinner's theory in which worldsheet correlators are manifestly permutation symmetric in the external states.  The worldsheet correlator itself in this formalism is given by:
\be{WSC}
\left\la \prod_{i=1}^{n}V^{h}_{i}\;\prod_{j=1}^{n-d-2+2g}\widetilde{\Upsilon}_{j}\;\prod_{k=1}^{d}\Upsilon_{d}\right\ra,
\ee
where the vertex operators are inserted at $\sigma_{i}\in\Sigma$, the $\beta\gamma$-system PCOs are inserted at $x_{j}\in\Sigma$, and the $\mu\nu$-system PCOs are inserted at $y_{k}\in\Sigma$.  The correlator is then computed by contracting all fields which have no zero-modes (e.g., $Y,\rho,\tilde{\rho}$) and then performing the remaining functional integrals over zero-modes.  As a check that this permutation-invariant setup is correct, let us compute the genus $g=0$ scattering amplitudes.

At genus zero $\Sigma\cong\P^{1}$, the line bundle $\cL\rightarrow\Sigma$ is simply $\cO(d)$, and the $\mathrm{m}\mathrm{n}$-system plays a trivial role,\footnote{There are no zero-modes of $\mathrm{m},\mathrm{m}_{ab}$, and the zero-modes of $\mathrm{n},\mathrm{n}_{ab}$ can be accounted for in the zero-mode integral for $Z$ \cite{Skinner:2013xp}.} so we are left to compute the amplitude
\begin{equation*}
 \int \frac{\prod_{a=0}^{d}\d^{4|8}U_{a}}{\mathrm{vol}\;\GL(2,\C)}\left\la \prod_{i=1}^{n}V^{h}(\sigma_i)\;\prod_{j=1}^{n-d-2}\widetilde{\Upsilon}(x_j)\;\prod_{k=1}^{d}\Upsilon(y_k)\right\ra,
\end{equation*}
where the $\{U_{a}\}$ are the parameters of the holomorphic map $Z:\P^{1}\rightarrow\PT$, providing a measure on its space of zero modes.  In Appendix \ref{GZFs}, we show that this worldsheet correlator leads to an expression for the amplitude:
\be{PMSA}
\cM_{n,d}=\int\frac{\prod_{a=0}^{d}\d^{4|8}U_{a}}{\mathrm{vol}\;\GL(2,\C)}\frac{|\mathsf{H}|}{|\mathsf{N}|^2}\frac{\left|\HH^{\vee}\right|}{|y_{1}\cdots y_{d}|^2}\prod_{i=1}^{n}h_i.
\ee

Here, the $n\times n$ matrix $\mathsf{N}$ is generated by the $\beta\gamma$-system, with entries:
\be{Nmat}
\mathsf{N}=\left( 
\begin{array}{cccccc}
\mathcal{Y}_{1}(\sigma_1) & \cdots & \mathcal{Y}_{d+2}(\sigma_1) & S(\sigma_{1},x_{1}) & \cdots & S(\sigma_{1},x_{n-d-2}) \\
\vdots & & \vdots & \vdots & & \vdots \\
\mathcal{Y}_{1}(\sigma_{n}) & \cdots & \mathcal{Y}_{d+2}(\sigma_{n}) & S(\sigma_{n},x_{1}) & \cdots & S(\sigma_{n}, x_{n-d-2}) 
\end{array}\right),
\ee 
with the $\mathcal{Y}_j$ a basis of zero modes for $\gamma_{a}$ and $S(\sigma_i,x_j)$ the propagator of the $\beta\gamma$-system.\footnote{There is a slight subtlety associated with the $\beta\gamma$-system propagator since these fields are charged under $\cL$; this has no meaningful impact on our claims and is discussed in detail in Appendix \ref{GZFs}.} Its determinant (known as a \emph{Slater determinant} \cite{Witten:2012bh}) appears in the denominator because the $\beta\gamma$-system is bosonic; the result is squared because the ghosts have two components.

The factor 
\begin{equation*}
 \frac{\left|\HH^{\vee}\right|}{|y_{1}\cdots y_{d}|^2},
\end{equation*}
is generated by the PCOs $\Upsilon_k$ of the $\mu\nu$-system, where $\HH^{\vee}$ is a $d\times d$ matrix known as the `dual Hodges matrix' \cite{Cachazo:2012kg} with entries
\be{DHM}
\HH^{\vee}_{kl}=\frac{\la Z(y_k),Z(y_l)\ra}{(y_{k}y_{l})} \;\;\mbox{for}\;k\neq l, \qquad \HH^{\vee}_{kk}=-\frac{\la Z(y_{k}),\partial Z(y_{k})\ra}{\D y_{k}}.
\ee
Finally, $\mathsf{H}$ is a $(n-d-2)\times(n-d-2)$-matrix with off-diagonal entries
\be{HM1}
\mathsf{H}_{jk}=\sum_{i=1}^{n}\sum_{l\neq i}\frac{\D x_{j}^{3/2}\;\D x_{k}^{3/2}}{(x_{j}x_{k})(x_{j}\sigma_{i})(x_{k}\sigma_{l})}\prod_{r=1}^{d+1}\frac{(a_{r}\sigma_{i})(a_{r}\sigma_{l})}{(a_{r}x_{j})(a_{r}x_{k})}\left[\frac{\partial}{\partial Z(\sigma_{i})},\frac{\partial}{\partial Z(\sigma_{l})}\right],
\ee
and diagonal entries
\be{HM2}
\mathsf{H}_{jj}=\sum_{i=1}^{n}\frac{\D x_j^{3}}{(x_{j}\sigma_{i})^2}\prod_{r=1}^{d+1}\frac{(a_{r}\sigma_{i})^2}{(a_{r}x_{j})^2}\sum_{l\neq i}\frac{1}{(\sigma_{i}\sigma_{l})}\prod_{s=1}^{d+1}\frac{(a_{s}\sigma_{l})}{(a_{s}\sigma_{i})}\left[\frac{\partial}{\partial Z(\sigma_{i})},\frac{\partial}{\partial Z(\sigma_{l})}\right].
\ee

While the expression for the genus zero S-matrix appears quite different from the Cachazo-Skinner formula \eqref{CSi}, the two are actually equivalent!  We derive \eqref{PMSA} explicitly and prove its equivalence with the Cachazo-Skinner formula in Appendix \ref{GZFs}.


\section{Worldsheet Gravity and Factorization}
\label{WSG}

In this section, we describe the main focus of this paper: worldsheet factorization.  Having set up a manifestly permutation symmetric formalism (at the level of external vertex operators) on the worldsheet, we now want to study how scattering amplitudes behave under multiparticle factorization.  In order to study this from the perspective of the worldsheet, we will need the $bc$-ghost system associated to worldsheet gravity.\footnote{Both the Berkovits-Witten and Skinner theories are chiral, so we will only need to include the holomorphic $bc$-system in each case.  Hence our discussion is analogous to the chiral sector of ordinary closed string theory.}  Of the theories we have described, only one (the Berkovits-Witten twistor-string) contains these Virasoro ghosts in the usual fashion.  In this section, we first discuss how to build in worldsheet gravity for each theory and then discuss the process of worldsheet factorization for a general theory at arbitrary worldsheet genus.


\subsection{Adding worldsheet gravity}

Worldsheet gravity, which is gauge-fixed using the Virasoro $bc$-ghost system, will be crucial for obtaining a clean description of worldsheet factorization in any string theory.  Let us begin by discussing the Berkovits-Witten twistor-string, which includes the $bc$-system in an anomaly-free way.  Recall that the vertex operators of this theory are given by $V^a, V^f, V^g$ from \eqref{BWVa}-\eqref{BWVfg}.  To study factorization, we want to fix the locations of all external states in the worldsheet correlator; integration over location then emerges as a moduli integral.  This can be accomplished by inserting a chiral `puncture operator' $\cP=c$ for each external state.  This gives us the set of fixed vertex operators
\be{BWfixed}
\cV^{a}=\oint c\;\tr(a(Z)\wedge j), \qquad \cV^f= \oint c\;Y_{I}\wedge f^{I}(Z), \qquad \cV^{g}=\oint c\;g_{I}(Z)\wedge\partial Z^{I},
\ee
integrating out the $(0,1)$-form component of each operator (since the theory is chiral) to leave a homogeneous scalar on the worldsheet $\Sigma$.  To recover the holomorphic integral over the worldsheet, we consider the pairing between a fixed vertex operator $\cV$ and an insertion of
\be{moduli}
\left(b|\mu\right)\equiv\int_{\Sigma}b\wedge\mu,
\ee
where $\mu\in\Omega^{0,1}(\Sigma,T_{\Sigma})$ is a Beltrami differential. These insertions correspond to tangent vectors on the worldsheet moduli space, and we must insert $3g-3+n$ to build a top degree form on the moduli space $\overline{M}_{g,n}$. 

To see this explicitly, work in local coordinates $(z,\bar{z})$ on $\Sigma$, where
\begin{equation*}
\left(b|\mu\right)=\int_{\Sigma}\d z\;\d\bar{z}\;\sqrt{g}\;b^{\alpha\beta}\D_{\alpha}v_{\beta},
\end{equation*}
where $v$ is a section of $T_{\Sigma}$.  The Wick contraction with a fixed vertex operator then gives
\begin{equation*}
\left\la (b(z)|\mu)\;c(z')\right\ra \propto\int_{\Sigma}\d z\;\d\bar{z}\delta(z-z'),
\end{equation*}
since the worldsheet derivative in $(b|\mu)$ acts on the standard two-point function $(z-z')^{-1}$ for the $bc$-system.  Hence, we conclude that
\begin{equation*}
\left\la (b|\mu)\;\cV\right\ra_{bc}=V,
\end{equation*}
with the worldsheet integral over $z'$ corresponding to the moduli integral with respect to the form $(b|\mu)$.  A more detailed discussion can be found in section 2.5 of \cite{Witten:2012bh}.

In this framework the gauge theory scattering amplitudes of the Berkovits-Witten twistor-string are given by:
\be{BWwsc}
\cA^{(g)}_{n,d}=\int\frac{\prod_{a=0}^{d-g}\d^{4|4}U_{a}}{\mathrm{vol}\; \C^{*}}\left\la\prod_{\alpha=1}^{3g-3+n}(b_{\alpha}|\mu_{\alpha})\;\prod_{i=1}^{n}\cV^{a}_{i} \right\ra.
\ee
When $g=0$, this is clearly equal to the RSV formula \eqref{RSV} since
\begin{equation*}
\left\la \prod_{\alpha=1}^{n-3}(b_{\alpha}|\mu_{\alpha})\;\prod_{i=1}^{n}\cV^{a}_{i}\right\ra_{bc}=\left\la \cV^{a}_{1}\cV^{a}_{2}\cV^{a}_{3}\right\ra_{bc}\prod_{j=4}^{n}V^{a}_{j}
=\frac{1}{\mathrm{vol}\;\SL(2,\C)}\prod_{i=1}^{n}V^{a}_{i}.
\end{equation*}

\medskip

Now, let us turn to Skinner's twistor-string, where the usual $bc$-ghost system is absent, leaving us without a natural set of coordinates for the (bosonic) worldsheet moduli.  It is in this sense that Skinner's theory is not actually a \emph{string theory}: at genus zero the $\SL(2,\C)$ automorphism group of $\Sigma\cong\P^1$--which is usually fixed by the 3 $c$-ghost insertions left over after Wick contracting with all $(b|\mu)$ insertions--must be fixed `by hand.'  So even at genus zero, we need some mechanism for describing worldsheet gravity if we are going to understand the factorization properties of the theory.

We proceed by taking a rather un-elegant route: simply add the $bc$-ghost action to the worldsheet theory of Skinner's model in \eqref{Sgh}.   Since this system doesn't couple to $\cL\rightarrow\Sigma$, the vanishing of the anomalies $\mathfrak{a}_{\GL(1)}$, $\mathfrak{a}_{\SL(2)}$, and $\mathfrak{a}_{\GL(1)/\mathrm{grav}}$ is undisturbed.  Unfortunately, the same cannot be said for the central charge of the theory, which is now $\mathfrak{c}=-26$.

This non-vanishing central charge obviously leads to problems in calculating general scattering amplitudes, or indeed interpreting the theory as a string theory at all.  In particular, the path integral over non-zero-modes of the $bc$-system will result in a factor of the determinant $\mathrm{det}'\dbar_{T_\Sigma}$, which must be interpreted as a section of some determinant line bundle over the moduli space $\overline{M}_{g}$ of genus $g$ Riemann surfaces:
\begin{equation*}
\mathrm{det}'\dbar_{T_{\Sigma}}\in\Gamma(\mathbb{L}), \qquad \xymatrix{\mathbb{L} \ar[r] & \overline{M}_{g} } .
\end{equation*}
For higher genus, the topology of this bundle could be quite complicated, and the factor $\mathrm{det}'\dbar_{T_{\Sigma}}$ may introduce new dependence on the moduli coordinates into the path integral.  However, for genus zero, $\overline{M}_{0}$ is just a point, so $\mathrm{det}'\dbar_{T_{\Sigma}}$ can just be treated as an overall numerical factor with respect to the worldsheet moduli.  While this number may affect the actual value of a scattering amplitude computed in the theory, it is irrelevant from the point of view of studying the moduli dependence of the worldsheet correlation function near a factorization limit.
 
Hence, we simply work with this anomalous modification of Skinner's theory.  As in the Berkovits-Witten theory, we can now write the vertex operators \eqref{VO} in a fixed representation:
\be{Sfixed}
\cV^{h}=\oint c\;\delta^{2}(\gamma)\;h(Z),
\ee
and the genus $g$ scattering amplitudes are computed by the worldsheet correlation function
\be{Swsc}
\cM_{n,d}^{(g)}=\int\frac{\prod_{a=0}^{d-g}\d^{4|8}U_{a}}{\mathrm{vol}\;\C^{*}}\left\la\prod_{\alpha=1}^{3g-3+n}(b_{\alpha}|\mu_{\alpha}) \prod_{i=1}^{n}\cV^{h}_i \prod_{j=1}^{n-d-2+2g}\widetilde{\Upsilon}_{j} \prod_{k=1}^{d}\Upsilon_{k}\right\ra.
\ee
At genus zero, this is easily seen to reduce to \eqref{WSC} (up to an overall factor of $\mathrm{det}'\dbar_{T_{\Sigma}}$) upon computing the parts of the correlator corresponding to the $bc$-system.

\medskip

Before proceeding, it is worth asking if there is some way for us to introduce worldsheet gravity into the twistor-string which does not modify the central charge.  Recall that in any two-dimensional CFT, the two-point function of the stress energy tensor is related to the central charge by
\begin{equation*}  
\left\la T(z)\;T(w)\right\ra = \frac{\mathfrak{c}/2}{(z-w)^4}+\cdots.
\end{equation*}
If the theory has a $b$-ghost which obeys $\{Q,b\}=T$ (for $Q$ the BRST charge) then the central charge automatically vanishes, since
\begin{equation*}
\left\la T(z)\;T(w)\right\ra=\left\la T(z)\;\{Q,b(w)\}\right\ra=\left\la\{Q,T(z)b(w)\}\right\ra =0,
\end{equation*}
via the BRST exactness of the stress-energy tensor.

Before inserting the standard $bc$-ghost system into Skinner's theory, the central charge was $\mathfrak{c}=0$.  If we could find some combination of ghost and matter fields $\mathcal{G}$ (from the unmodified twistor-string) which obeyed $\{Q,\mathcal{G}\}=T$, then it would define an effective $b$-ghost: $b^{\mathrm{eff}}=\mathcal{G}$.  Clearly whatever combination works must be a fermionic quadratic differential on the worldsheet $\Sigma$, so the measure on the moduli space $\overline{M}_{g,n}$ could then be built by inserting
\begin{equation*}
 \prod_{\alpha=1}^{3g-3+n}\left(b^{\mathrm{eff}}_{\alpha}|\mu_{\alpha}\right),
\end{equation*}
in the path integral.  Crucially, the central charge would remain $\mathfrak{c}=0$, since this procedure introduces no new fields in the worldsheet action. 

Unfortunately, no such effective $b$-ghost for the twistor-string is currently known.  One could hope to take a cue from the pure spinor formalism \cite{Berkovits:2000fe}, where similar issues arise.  There the worldsheet action also lacks the conventional $bc$-ghost system, and the effective ghost $b^{\mathrm{eff}}$ is built from a rather complicated combination of fields \cite{Berkovits:2004px, Berkovits:2005bt} engineered in such a way that $\{Q,b^{\mathrm{eff}}\}=T$.\footnote{Here we have in mind the `non-minimal' pure spinor formalism, in which $b^{\mathrm{eff}}$ only has potential poles of the form $(\bar{\lambda}\lambda)^{-k}$ for $k=0,\ldots,4$.}  In this construction it is crucial that there are fields (namely, the pure spinor itself) which are BRST invariant.  However there are no such fields in the twistor-string, which makes it seem unlikely that an effective $b$-ghost can be built using a similar strategy.  We discuss this issue further in Section \ref{Concl}.


\subsection{The factorization limit}
\label{FL}

We conclude this section with an overview of worldsheet factorization; we are rather brief here, and our treatment follows that of section 6 in \cite{Witten:2012bh}.  In momentum space, scattering amplitudes have poles corresponding to internal propagators going on-shell.  For instance, in a $n$-particle scattering amplitude with external momenta $\{p_{i}\}$, we can consider the limit where a subset of the momenta go on-shell: $P_{L}^{2}=(\sum_{i=1}^{n_L}p_i)^{2} =0$.  Unitarity demands that the scattering amplitude obeys:
\begin{equation*}
\cM(p_{1},\ldots,p_{n})\rightarrow \delta^{4}\left(\sum_{i=1}^{n}p_i\right)\int \d^{\cN}\eta\;\cM^{L}(p_{1},\ldots,p_{n_L},P_{L})\frac{1}{P_L^{2}}\cM^{R}(-P_{L},p_{n_L+1},\ldots,p_{n}),
\end{equation*}
where $P_{L}$ is the momentum of the internal propagator which is going on-shell, and the supersymmetric integral accounts for a sum over relevant helicity configurations.  The residue of the simple pole in $P_{L}^{2}$ factorizes the original amplitude into two new amplitudes $\cM^{L}$, $\cM^{R}$ of on-shell states.  It is well known that these multiparticle factorization poles correspond to the IR region of the field theory under consideration.

In string theory, the analogue of an internal propagator going on-shell is a singularity in the worldsheet topology which `pinches' the worldsheet.  Approaching this limit is conformally equivalent to considering a worldsheet which develops a long, thin tube that is eventually pinched shut.  In the Deligne-Mumford compactification of the moduli space of Riemann surfaces, this limit simply corresponds to approaching a boundary divisor.  The role of the massless propagator $p^{-2}$ in field theory is played by the string propagator (c.f., \cite{Green:1987}, Chapter 7 or \cite{Polchinski:1998}, Chapter 9), which can in turn be viewed as the measure on the worldsheet moduli space near the boundary divisor (c.f., \cite{Polchinski:1988jq, Witten:2012bh}).

Hence, investigating factorization from this worldsheet perspective has the substantial advantage of being intrinsically geometric.  Demonstrating that scattering amplitudes have the appropriate factorization properties is reduced to showing that as we approach the boundary divisor, the worldsheet correlator has the structure of a $p^{-2}$ pole in momentum space.  In this overview, we will consider the moduli space of (bosonic) Riemann surfaces with the factorization limit corresponding to a boundary divisor $\mathfrak{D}\subset\overline{M}_{g,n}$, akin to bosonic string theory.

A generic boundary divisor $\mathfrak{D}\subset\overline{M}_{g,n}$ will either take the form of a \emph{separating} or \emph{non-separating} divisor (c.f., \cite{Witten:2012bh}):
\begin{equation*}
\mathfrak{D}^{\mathrm{sep}}\cong\overline{M}_{g_{L},n_{L}+1}\times \overline{M}_{g_{R},n_{R}+1}, \qquad \mathfrak{D}^{\mathrm{ns}}\cong\overline{M}_{g-1,n+2},
\end{equation*}
respectively.  Here $g_{L}+g_{R}=g$ and $n_{L}+n_{R}=n$ in the separating case; one can show that in both cases $\mathfrak{D}$ has codimension one with respect to $\overline{M}_{g,n}$.

For both degenerations of the worldsheet, we want to model the geometry of $\Sigma$ near the divisor and in the neighborhood of the singularity.  This can be accomplished by gluing together two Riemann surfaces $\Sigma_{L}$, $\Sigma_{R}$ along a tube.  In the separating case, $\Sigma_{L}$ and $\Sigma_{R}$ can be thought of as the two surfaces resulting from the singular limit; in the non-separating case we simply think of them as providing a local image of the worldsheet in the neighborhood of a degenerating cycle on $\Sigma$.

\begin{figure}[t]
\centering
\includegraphics[width=2.5 in, height=2.0 in]{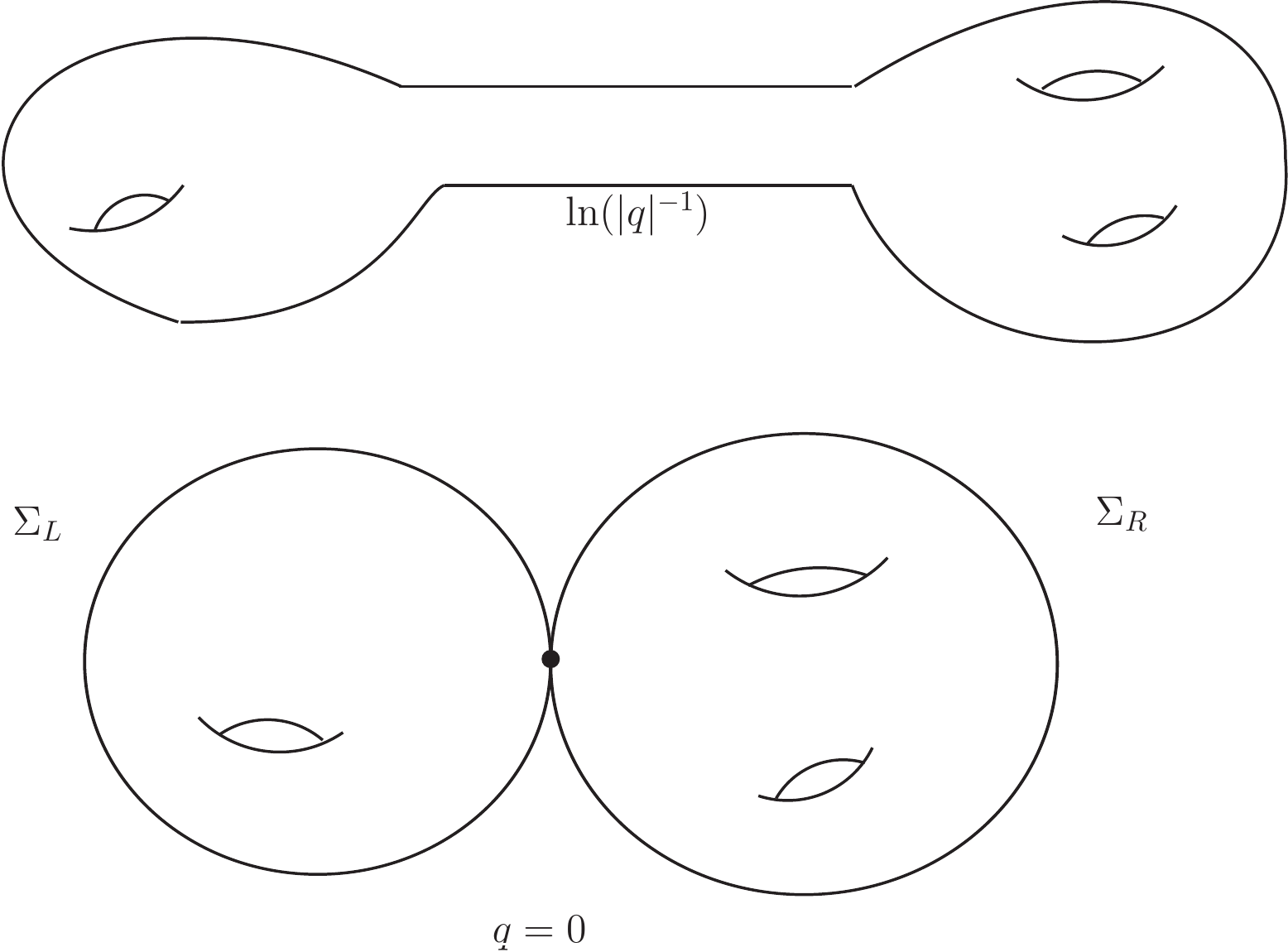}\caption{\small{\textit{A genus three worldsheet approaches the separating divisor $\mathfrak{D}^{\mathrm{sep}}$ as the modulus $q$ approaches zero.}}}\label{TSFac1}
\end{figure}

\begin{figure}[t]
\centering
\includegraphics[width=2.5 in, height=1.35 in]{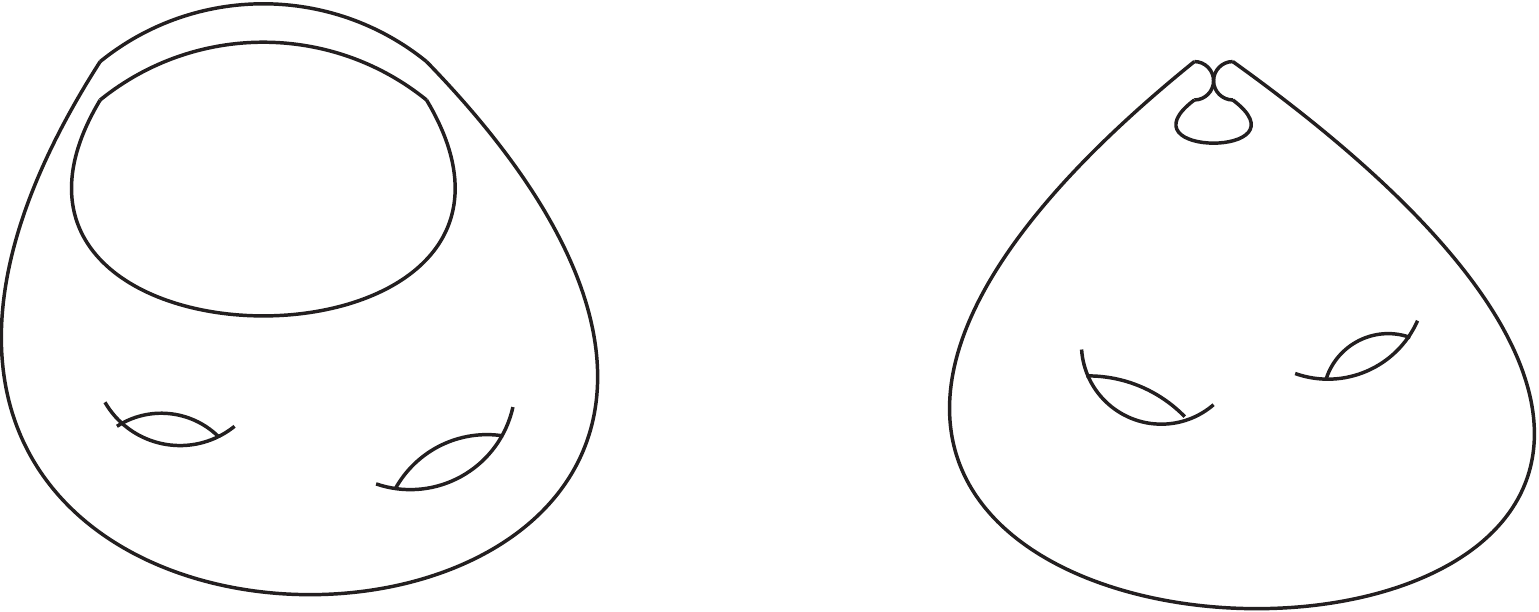}\caption{\small{\textit{The non-separating degeneration reduces the genus of the worldsheet by one.}}}\label{NSFac}
\end{figure}

Let $z_{i}$ be a local coordinate on $\Sigma_{i}$, for $i=L,R$.  Our local model is given by
\be{WF1}
(z_{L}-a)(z_{R}-b)=q,
\ee
which for $q=0$ has two branches: $\Sigma_{L}=\{z_{R}=b\}$ and $\Sigma_{R}=\{z_{L}=a\}$.  This models the singular limit by attaching the point $a\in\Sigma_{L}$ to the point $b\in\Sigma_{R}$.  Near $q=0$, \eqref{WF1} locally describes $\Sigma_{L}$ and $\Sigma_{R}$ joined by a neck of modulus $q$.  To be precise, the length of this neck is given by $\ln(|q|^{-1})$, so the geometry as $q\rightarrow0$ is conformally equivalent to joining the two $q=0$ branches of \eqref{WF1}.  Figure \ref{TSFac1} illustrates this in the case of the separating divisor, while Figure \ref{NSFac} shows the non-separating degeneration.

For the purposes of worldsheet factorization we are interested in what the measure on the moduli space looks like near the boundary divisor where $q=0$.  Let us discuss this in detail for the separating divisor $\mathfrak{D}^{\mathrm{sep}}$; the basic ideas carry over to the non-separating case.  Near the boundary, the $3g-3+n$ moduli of the worldsheet $\Sigma$ can be decomposed into moduli $\mathsf{x}_{\alpha}$ on $\Sigma_{L}$ (for $\alpha=1,\ldots,3g_{L}-3+n_L$) and $\mathsf{y}_{\beta}$ on $\Sigma_{R}$ (for $\beta=1,\ldots,3g_{R}-3+n_2$).  This leaves an apparent deficit of three moduli, which are accounted for by the addition of $\mathsf{x}_{a}$ (for the position of the point $a\in\Sigma_{L}$), $\mathsf{y}_{b}$ (for the position of $b\in\Sigma_{R}$), and the modulus $q$ itself, which can be thought of as a coordinate transverse to the boundary divisor $\mathfrak{D}^{\mathrm{sep}}\subset\overline{M}_{g,n}$.

Hence, near the boundary divisor, the measure on the moduli space takes the form
\be{WF2}
\prod_{\alpha=1}^{3g_{L}-3+n_L}\d\mathsf{x}_{\alpha} \prod_{\beta=1}^{3g_{R}-3+n_{R}}\d\mathsf{y}_{\beta}\;f(q)\;\d\mathsf{x}_{a}\;\d\mathsf{y}_{b}\;\d q,
\ee
where $f(q)$ is an overall factor whose $q$-dependence is fixed by homogeneity.  Due to the local nature of \eqref{WF1}, it is important to keep in mind that this form of the measure on the moduli space only makes sense when $|q|$ is small, or we are near the boundary divisor.  As $q\rightarrow0$, we get a sum over states inserted at $a\in\Sigma_L$ and $b\in\Sigma_R$; this is a basic property of the worldsheet theory which is inherited from the OPE of two-dimensional CFT (c.f., \cite{Vafa:1987ea, Polchinski:1988jq} or \cite{Polchinski:1998} Chapter 9 for a review). The form $f(q)\d q$ gives the corresponding string propagator between these states.  The scattering amplitudes of the string theory factorize unitarily at the level of the worldsheet provided the insertion of on-shell physical vertex operators corresponds to a simple pole of the form $q^{-1}\d q$.

As an example, consider the insertion of identity operators on each side of the singularity.  Then the portion of the moduli measure \eqref{WF2} which we are interested in looks like
\begin{equation*}
f(q)\;\d\mathsf{x}_{a}\;\d\mathsf{y}_{b}\;\d q. 
\end{equation*}
This measure must be homogeneous under scalings of the individual moduli.  From \eqref{WF1}, we see that for a re-scaling $\mathsf{x}_{a}\rightarrow\lambda\mathsf{x}_{a}$, $\mathsf{y}_{b}\rightarrow\tilde{\lambda}\mathsf{y}_{b}$, the variable $q$ must scale as $q\rightarrow\lambda\tilde{\lambda}q$.  Homogeneity then dictates that $f(q)=q^{-2}$, leaving us with the measure
\begin{equation*}
\d\mathsf{x}_{a}\;\d\mathsf{y}_{b}\;\frac{\d q}{q^2},
\end{equation*}
which does \emph{not} have a simple pole in $q$.  Hence, the identity operator cannot contribute to the pole relevant for factorization of the string theory scattering amplitudes.


\section{Genus Zero Factorization}
\label{GZero}

Studying worldsheet factorization in twistor-string theories corresponds to testing the behavior of their scattering amplitudes under multiparticle factorization.  At tree level, the RSV and Cachazo-Skinner formulae are known to factorize correctly; combined with the correct 3-point amplitudes and tests of soft and collinear limits, this \emph{proves} that the formulae for the amplitudes are correct.  However, actually testing factorization at the level of the amplitude itself is a rather cumbersome task: scalings in the factorization limit are buried in the details of the Wick contractions which have produced the answer, and isolating the correct pole structure can be difficult.

In this section, we re-derive the correct multiparticle factorization behavior for all tree-amplitude formulae by considering the genus zero worldsheet factorization of their respective twistor-string theories.  In the Berkovits-Witten twistor-string unitary factorization for gauge theory amplitudes follows in the planar limit, as demonstrated at the level of the answer by \cite{Skinner:2010cz, Dolan:2011za}.  However, when all physical states are allowed to propagate (i.e., away from the planar limit) we discover a double pole which is consistent with the appearance of conformal supergravity in the theory \cite{Berkovits:2004jj}.  For the Skinner theory we show that only the simple pole corresponding to the correct factorization behavior demonstrated in \cite{Cachazo:2012pz} appears.  In all cases, the derivation of the factorization properties at the level of the worldsheet is significantly easier than that given by studying the amplitude itself.

We then contrast factorization in twistor-string theory with the situation in ordinary string theory.  As we will see, the main difference is in the interpretation of the moduli coordinate transverse to the boundary divisor.  In the twistor-string, choosing momentum eigenstates for the external vertex operators allows us to directly identify powers of $q$ with $p^{2}$ factors for an internal propagator; in ordinary string theory this is only true for a simple pole in $q$.  The difference is due to the fact that all states in twistor-string theory are conformal weight zero, and specified by a generic cohomology class.  


\subsection{Berkovits-Witten and conformal supergravity}

In this paper we consider only the gauge theoretic scattering amplitudes of the Berkovits-Witten twistor-string.  Explicitly incorporating worldsheet gravity, these are given by the genus zero form of \eqref{BWwsc}: 
\be{BWgz1}
\cA^{(0)}_{n,d}=\int\frac{\prod_{a=0}^{d}\d^{4|4}U_{a}}{\mathrm{vol}\; \C^{*}}\left\la\prod_{\alpha=1}^{n-3}(b_{\alpha}|\mu_{\alpha})\;\prod_{i=1}^{n}\cV^{a}_{i} \right\ra.
\ee
Here the moduli space being integrated over is $\overline{M}_{0,n}(\PT,d)$: the moduli space of holomorphic maps from $\Sigma\cong\P^1$ to twistor space $\PT$ of degree $d$.  Since twistor space is a projective supermanifold, this moduli space can be understood as a natural super-geometric generalization of Kontsevich's moduli space of stable maps \cite{Adamo:2012cd}.  In particular, this means that it inherits its boundary structure entirely from the compactification of $\overline{M}_{0,n}$, where the only boundary divisors correspond to the separating factorization channel:
\begin{equation*}
\mathfrak{D}^{\mathrm{sep}}\cong\overline{M}_{0,n_{L}+1}(\PT,d_{L})\times \overline{M}_{0,n_{R}+1}(\PT,d_{R}),
\end{equation*}
with $d_{L}+d_{R}=d$.

As discussed in the previous section, the appropriate local model for worldsheet factorization is given by the divisor in $\mathfrak{D}^{\mathrm{sep}}\subset\overline{M}_{0,n}$.  We want to study the behavior of $\cA^{(0)}_{n,d}$ near $\mathfrak{D}^{\mathrm{sep}}$ using the local model \eqref{WF1} for the worldsheet.  Suppose the vertex operators $\cV^{a}_{i}$ distribute themselves arbitrarily over the two factors in a way compatible with the cyclic ordering of the overall gauge theoretic trace, with $n_{L}$ on $\Sigma_{L}$ and $n_{R}$ on $\Sigma_{R}$.  Clearly, $n_{L}+n_{R}=n$, and we assume for convenience that $n_{L},n_{R}\geq2$.\footnote{The subtleties associated with the insertion of fewer than two vertex operators on one of the factors correspond to mass renormalization and massless tadpoles in string perturbation theory.  A discussion of these issues can be found in Section 7 of \cite{Witten:2012bh}, and we will gloss over them for the remainder of this paper.}

The forms $(b_{\alpha}|\mu_{\alpha})$ also distribute themselves among the two factors.  While they can do so in an arbitrary fashion, it is not hard to see that the resulting amplitude will be zero unless there are $n_{L}-3$ on $\Sigma_{L}$ and $n_{R}-3$ on $\Sigma_{R}$, with the remaining three corresponding to the moduli $\mathsf{x}_{a}$, $\mathsf{y}_{b}$, and $q$ in the measure \eqref{WF2}.\footnote{Any other distribution of the $(b_{\alpha}|\mu_{\alpha})$ would not result in a top-degree form on the moduli space for $\Sigma_{L}$ or $\Sigma_{R}$.}  Hence, as we approach $\mathfrak{D}^{\mathrm{sep}}$ (i.e., the $q\rightarrow 0$ limit) the worldsheet correlator takes the form:
\be{BWgz2}
\prod_{\alpha=1}^{n_{L}-3}(b_{\alpha}|\mu_{\alpha})\prod_{i=1}^{n_{L}}\cV_{i}^{a}\:\prod_{\beta=1}^{n_{R}-3}(b_{\beta}|\mu_{\beta}) \prod_{j=1}^{n_{R}}\cV_{j}^{a} \sum_{\mathrm{states}}(b_{a}|\mu_{a})\cO_{a}\;f(q)\;\d q\;(b_{b}|\mu_{b})\cO_{b},
\ee
where the sum is over states $\cO$ in the twistor-string which propagate across the cut in the worldsheet.  State $\cO_{a}$ is inserted at $a\in\Sigma_{L}$, the anti-state $\cO_{b}$ is inserted at $b\in\Sigma_{R}$, and the scaling function $f(q)$ is fixed by homogeneity of the measure on the moduli space. 

\begin{figure}[t]
\centering
\includegraphics[width=4.5 in, height=0.3 in]{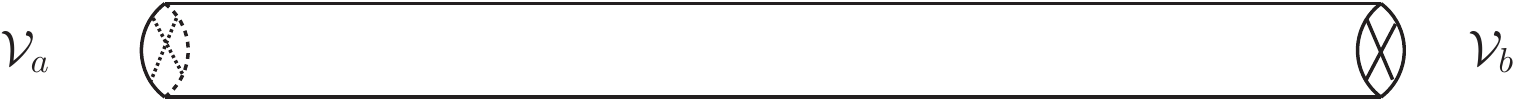}\caption{\small{\textit{Near the boundary divisor, $\cV_{a}$ and $\cV_{b}$ are inserted at opposite ends of a long tube.}}}\label{Tube}
\end{figure}

In the Berkovits-Witten twistor-string, physical states are given by $\cV^{a}$, $\cV^{f}$, or $\cV^{g}$.  Hence, we consider terms in \eqref{BWgz2} for which $\cO_{a,b}=\cV_{a,b}$, where $\cV_{a,b}$ are one of the physical vertex operators.  Very close to the boundary divisor, this looks like inserting $\cV_{a}$ and $\cV_{b}$ at opposite ends of a very long tube joining $\Sigma_L$ to $\Sigma_R$, as illustrated in Figure \ref{Tube}.  We can actually restrict which of these vertex operators appear in the factorization channel by taking a closer look at the structure of the worldsheet correlation function.

Notice that in the initial correlator, the only matter system in play at the level of the external vertex operators is the worldsheet current algebra.  The twistor wavefunctions $a(Z)$ are all functions of $Z$, which cannot Wick contract with anything else in the correlator and are therefore simply integrated over the space of zero modes (i.e., holomorphic maps).  Hence, we can classify different contributions to the scattering amplitude by looking at the correlator in the worldsheet current algebra.

Suppose we take the following explicit representation for the action of the worldsheet current algebra:
\begin{equation*}
S_{C}[\lambda]=\int_{\Sigma}\tr\left(\bar{\lambda}\wedge\dbar\lambda\right), \qquad \lambda\in\Pi\Omega^{0}(\Sigma, K^{\frac{1}{2}}_{\Sigma}\otimes\mathfrak{g}), 
\end{equation*}
where $\mathfrak{g}$ is the Lie algebra of some gauge group (which we assume to be traceless for convenience).  The explicit form of the current appearing in the vertex operators is
\be{current}
j^{a}(\sigma)=\mathsf{T}^{ai}_{\;j}\;\bar{\lambda}_{i}(\sigma)\;\lambda^{j}(\sigma),
\ee
where $\mathsf{T}^{a}$ are the generators of the fundamental representation of $\mathfrak{g}$.  Clearly, the genus zero scattering amplitudes will be proportional to traces over these generators.  We can classify contributions to the worldsheet correlator by the number of such traces: single trace, double trace, and so forth.  Each of these will be proportional to factors of the form
\begin{equation*}
 \tr\left(\mathsf{T}^{a_1}\cdots\mathsf{T}^{a_n}\right), \qquad \tr(\mathsf{T}^{a_1}\cdots\mathsf{T}^{a_{n_1}})\;\tr(\mathsf{T}^{a_{n_1+1}}\cdots\mathsf{T}^{a_n}), 
\end{equation*}
respectively.

In the planar limit for the gauge algebra $\mathfrak{g}$, only single trace contributions to the worldsheet correlator appear, as all multi-trace terms are suppressed.  For the purposes of worldsheet factorization, this means that only gauge theoretic vertex operators can appear in the factorization channel.  So in the planar limit, we can take $\cO_{a,b}=\cV^{a}_{a,b}$ in \eqref{BWgz2}.  

Now we need to fix the scaling function $f(q)$.  To do this, consider the behavior of the measure under a rescaling $\mathsf{x}_{a}\rightarrow \lambda\mathsf{x}_{a}$ and $\mathsf{y}_{b}\rightarrow\tilde{\lambda}\mathsf{y}_{b}$.  Since $\cV^{a}$ is a BRST-closed vertex operator of conformal weight $(0,0)$, it follows that under $\mathsf{x}_{a}\rightarrow \lambda\mathsf{x}_{a}$
\begin{equation*}
(b_{a}|\mu_{a})\;\cV^{a}_{a}\xrightarrow{\mathsf{x}_{a}\rightarrow\lambda\mathsf{x}_{a}} (b_{a}|\mu_{a})\;\cV^{a}_{a},
\end{equation*}
and likewise for the insertions at $b\in\Sigma_{R}$.  However, the moduli coordinate $q$ still scales as $q\rightarrow\lambda\tilde{\lambda}q$ thanks to the local model for the worldsheet \eqref{WF1}.  In order for the measure on the moduli space to be homogeneous, we must have $f(q)=q^{-1}$ for the term in \eqref{BWgz2} with physical states.

\begin{figure}[t]
\centering
\includegraphics[width=5.5 in, height=1.2 in]{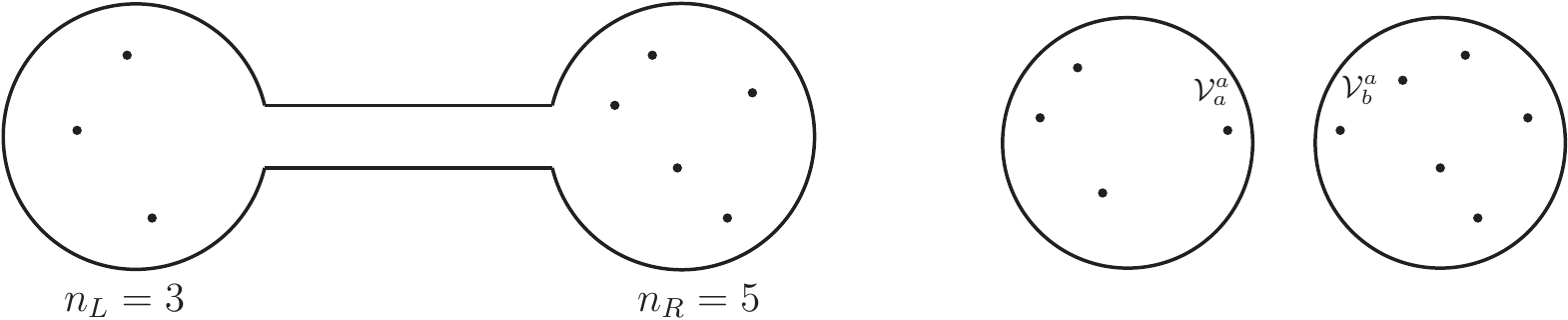}\caption{\small{\textit{Worldsheet factorization in the planar limit of the Berkovits-Witten twistor-string at 8-points.}}}\label{TSFac2}
\end{figure}

Near the boundary of the moduli space, this means that the worldsheet correlator takes the form: 
\be{BWgz3}
\prod_{\alpha=1}^{n_{L}-2}(b_{\alpha}|\mu_{\alpha})\prod_{i=1}^{n_{L}+1}\cV_{i}^{a}\;\prod_{\beta=1}^{n_{R}-2}(b_{\beta}|\mu_{\beta}) \prod_{j=1}^{n_{R}+1}\cV_{j}^{a} \;\frac{\d q}{q},
\ee
where the $\alpha=n_{L}-2$ and $i=n_{L}+1$ factors correspond to the new insertion at $a\in\Sigma_{L}$, and likewise for $b\in\Sigma_{R}$.  There is a simple pole in the modulus $q$, which can be interpreted as a momentum space propagator $p^{-2}$ (see Appendix \ref{QProps}).  As we approach the divisor $\mathfrak{D}^{\mathrm{sep}}$, potential Wick contractions between the factors $\Sigma_L$ and $\Sigma_R$ are suppressed as the tube connecting them becomes infinitely long (we demonstrate this explicitly in Appendix \ref{QProps}).  Hence, the residue of the $q=0$ pole in \eqref{BWgz3} has the structure of two on-shell worldsheet correlation functions (one on $\Sigma_{L}$ and one on $\Sigma_{R}$).  An example of an 8-point scattering amplitude is illustrated in Figure \ref{TSFac2}.

All that remains is to account for the map portion of the moduli.  These are encoded in the measure factor from \eqref{BWgz1}:
\begin{equation*}
\frac{\prod_{a=0}^{d}\d^{4|4}U_{a}}{\mathrm{vol}\; \C^{*}}.
\end{equation*}
Near the boundary divisor $\mathfrak{D}^{\mathrm{sep}}$, we write this measure as
\begin{equation*}
\d^{4|4}U_{\bullet} \prod_{a=1}^{d_{L}}\d^{4|4}U_{a} \prod_{b=1}^{d_{R}}\d^{4|4}W_{b},
\end{equation*}
where the node at $q=0$ is mapped to $U_{\bullet}\in\PT$.  Now, we insert into the path integral an inventive factor of unity:
\begin{equation*}
1=\int \D^{3|4}W_{\bullet}\;\D^{3|4}Z\;\bar{\delta}^{3|4}(Z,U_{\bullet})\;\bar{\delta}^{3|4}(Z,W_{\bullet}),
\end{equation*}
where $\bar{\delta}^{3|4}$ is a homogeneous $(0,3)$-form on $\PT$ enforcing the projective coincidence of its two arguments:
\begin{equation*}
 \bar{\delta}^{3|4}(Z,Y)\equiv \int_{\C}\frac{\d s}{s}\bar{\delta}^{4|4}(Z+s Y).
\end{equation*}

At this stage, we pull together all the pieces and take the residue of the worldsheet correlator at $q=0$.  At the level of the worldsheet fields, this factorizes the correlator in \eqref{BWgz3} into two separate correlators: one on $\Sigma_{L}$ and the other on $\Sigma_{R}$, which we denote in shorthand by
\begin{equation*}
 \left\la \Sigma_{L}\right\ra = \left\la \prod_{\alpha=1}^{n_{L}-2}(b_{\alpha}|\mu_{\alpha})\prod_{i=1}^{n_{L}+1}\cV_{i}^{a}\right\ra.
\end{equation*}
The amplitude $\cA^{(0)}_{n,d}$ then behaves as:
\begin{multline}\label{BWgz4}
\cA^{(0)}_{n,d}\xrightarrow{q=0} \int\frac{\d^{4|4}U_{\bullet}}{\mathrm{vol}\;\C^{*}} \prod_{a=1}^{d_{L}}\d^{4|4}U_{a} \prod_{b=1}^{d_{R}}\d^{4|4}W_{b} \left\la\Sigma_{L}\right\ra\;\left\la\Sigma_{R}\right\ra  \\
=\int \D^{3|4}W_{\bullet}\;\D^{3|4}Z\;\bar{\delta}^{3|4}(Z,U_{\bullet})\;\bar{\delta}^{3|4}(Z,W_{\bullet})\frac{\d^{4|4}U_{\bullet}}{\mathrm{vol}\;\C^{*}} \prod_{a=1}^{d_{L}}\d^{4|4}U_{a} \prod_{b=1}^{d_{R}}\d^{4|4}W_{b} \left\la\Sigma_{L}\right\ra\;\left\la\Sigma_{R}\right\ra \\
=\int \D^{3|4}Z \frac{\prod_{a=0}^{d_{L}}\d^{4|4}U_{a}}{\mathrm{vol}\;\C^{*}}\frac{\prod_{b=0}^{d_{R}}\d^{4|4}W_{b}}{\mathrm{vol}\;\C^{*}}\left\la\Sigma_{L}\right\ra\;\left\la\Sigma_{R}\right\ra =\int \D^{3|4}Z\;\cA^{(0)}_{n_{L}+1,d_{L}}\;\cA^{(0)}_{n_{R}+1,d_{R}}.
\end{multline}
Here, the delta functions $\bar{\delta}^{3|4}(Z,U_{\bullet})$ and $\bar{\delta}^{3|4}(Z,W_{\bullet})$ have been absorbed into the definition of the new external states in $\cV^{a}_{a}$ and $\cV^{a}_{b}$ respectively, since they enforce the insertion of these operators at the pinched node of the original worldsheet.

The final line of \eqref{BWgz4} is precisely the expression for multiparticle factorization on twistor space \cite{Mason:2009sa, Skinner:2010cz}.  This proves that the RSV formula (i.e., the leading trace gauge theory amplitudes of the Berkovits-Witten twistor-string) factorizes appropriately.  Although we have gone through the reasoning in detail, this follows immediately after obtaining the measure for the moduli space near the factorization limit in \eqref{BWgz3}.  Indeed, the simple pole in the moduli coordinate $q$ is the string theoretic analogue of the simple momentum pole $p^{-2}$ expected in field theory.  To arrive at \eqref{BWgz4}, we only needed to know about the geometry of the moduli space and the vertex operators of the theory.  No Wick contractions or integrals in the scattering amplitude were actually needed.  This highlights the applicability of worldsheet factorization, even at genus zero: it gives a substantially simpler proof of the IR properties of the scattering amplitudes than an investigation of the final answer itself.

\medskip

Let us now move away from the planar sector, and consider the multi-trace contributions to a $n$-point, degree $d$ scattering amplitude in the Berkovits-Witten theory.  We will focus on the double trace contribution, and the following discussion applies in the obvious way to the multi-trace generalizations.  Such a double trace contribution takes the form
\be{CGF1}
\cA^{\mathrm{DT}}_{n,d}=\tr(\mathsf{T}^{a_1}\cdots\mathsf{T}^{a_{n_L}})\;\tr(\mathsf{T}^{b_1}\cdots\mathsf{T}^{b_{n_R}})\int\frac{\prod_{a=0}^{d}\d^{4|4}U_{a}}{\mathrm{vol}\; \C^{*}}\left\la\prod_{\alpha=1}^{n-3}(b_{\alpha}|\mu_{\alpha})\;\prod_{i=1}^{n_{L}}\widetilde{\cV}_{a_i}\prod_{j=1}^{n_{R}}\widetilde{\cV}_{b_j} \right\ra,
\ee
where $n_L+n_R=n$ and $\widetilde{\cV}_{a_i}$ is shorthand for the vertex operator $\cV^{a}_i$ stripped of the generator $\mathsf{T}^{a_i}$ in accordance with \eqref{current}.

It has been known for nearly a decade that multi-trace contributions such as \eqref{CGF1} have factorization behavior consistent with conformal gravity degrees of freedom.  Indeed, this was first noted at the level of the four-point amplitude in \cite{Witten:2003nn}.  To show this at the level of the worldsheet, we consider a particular factorization channel of \eqref{CGF1}: the one that does not disturb the trace structure of the worldsheet current algebra.  As our notation suggests, this means that as we approach the separating boundary divisor $\mathfrak{D}^{\mathrm{sep}}$, the vertex operators $\{\widetilde{\cV}_{a_i}\}_{i=1,\ldots,n_L}$ are on $\Sigma_L$ while $\{\widetilde{\cV}_{b_j}\}_{j=1,\ldots,n_R}$ are on $\Sigma_R$.

The condition that we do nothing to disturb the color trace of the worldsheet current algebra means that we only focus on terms in the factorization limit where \emph{gravitational} degrees of freedom appear on either side of the cut.  This means that as we approach the boundary divisor, the worldsheet correlator is given by:
\begin{multline}\label{CGF2}
\tr(\mathsf{T}^{a_1}\cdots\mathsf{T}^{a_{n_L}})\; \tr(\mathsf{T}^{b_1}\cdots\mathsf{T}^{b_{n_R}}) \prod_{\alpha=1}^{n_L-3}(b_{\alpha}|\mu_{\alpha})\left\la\prod_{i=1}^{n_L}\widetilde{\cV}_{a_i}\right\ra_{S_C} \prod_{\beta=1}^{n_R-3}(b_{\beta}|\mu_{\beta})\left\la\prod_{j=1}^{n_R}\widetilde{\cV}_{b_j}\right\ra_{S_C} \\
\times (b_{a}|\mu_{a})\;\cV^{f}_{a}\;f(q)\d q\;(b_{b}|\mu_{b})\;\cV^{g}_{b},
\end{multline}
where $\la\cdots\ra_{S_C}$ denotes the correlation function with respect to the worldsheet current algebra, and $\cV^{f}$, $\cV^{g}$ are the gravitational vertex operators \eqref{BWVfg}.\footnote{One could also have terms in this factorization channel with two $\cV^{f}$s or two $\cV^{g}$s; these will have the same factorization behavior as the mixed contributions we have chosen.  We choose to work with \eqref{CGF2} as an explicit example since this is the only non-vanishing contribution at four points, which is the first non-trivial example of a double-trace contribution.  In particular, the only non-vanishing three-point factors of the double-trace term in $\la\cV^{a}_{1}\cV^{a}_{2}\cV^{a}_{3}\cV^{a}_{4}\ra_{d=1}$ are $\la\cV^{a}_{1}\cV^{a}_{2}\cV^{f}_{a}\ra_{d=0}$ and $\la\cV^{g}_{b}\cV^{a}_{3}\cV^{a}_{4}\ra_{d=1}$ \cite{Berkovits:2004jj}.}

As in the planar case, we now need to determine the scaling function $f(q)$ in \eqref{CGF2}.  Since both $\cV^{f}$ and $\cV^{g}$ have zero conformal weight, it may at first seem like we should have $f(q)=q^{-1}$.  However, upon inspection it is clear that \eqref{CGF2} exhibits a \emph{doubled} factorization channel.  This is due to the explicit form of the vertex operators $\cV^{a}$: since the wavefunctions $a(Z)$ do not Wick contract with each other (or any of the other insertions in \eqref{CGF1}), the correlator is effectively factorized before even approaching a boundary divisor.  By singling out the factorization channel which does not disturb the color trace, we ensure that the gauge theoretic structure of the amplitude is unchanged as the worldsheet degenerates.  Furthermore, the new gravitational vertex operators do not couple to the worldsheet current algebra, so this structure is preserved as we approach the boundary divisor.

\begin{figure}[t]
\centering
\includegraphics[width=3.8 in, height=1.2 in]{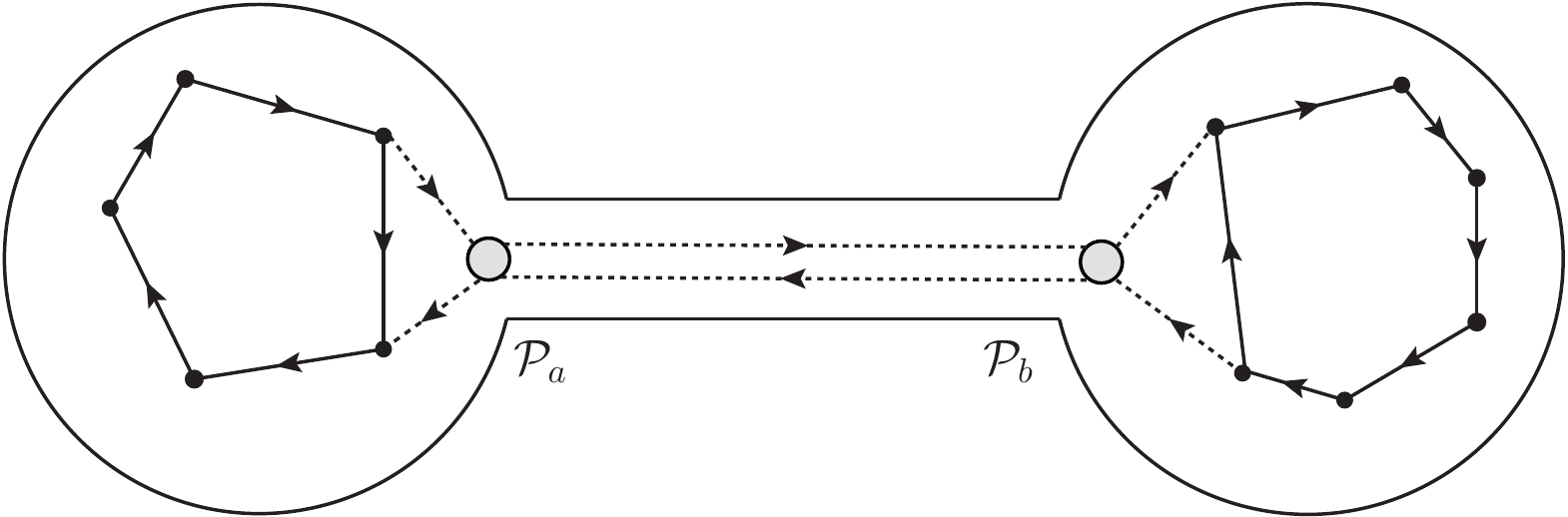}\caption{\small{\textit{The puncture operators effectively cut the single trace (dotted lines) into a double trace (solid lines) as we approach the boundary divisor.}}}\label{TSFac4}
\end{figure}

A more formal way of stating this is that the double trace contribution \eqref{CGF1} could be obtained by factorizing a single trace term, but with puncture operators $\cP=c$ inserted on each side of the cut rather than the gauge theory vertex operators appearing in \eqref{BWgz3}.  The role of these puncture operators is to cut the worldsheet correlator into two factors corresponding to each trace, as illustrated in Figure \ref{TSFac4}. The factorization channel in \eqref{CGF2} cuts the worldsheet in the same way, so the modulus $q$ is a coordinate transverse to \emph{both} of these cuts. 

This lets us re-write \eqref{CGF2} as 
\be{CGF*}
\prod_{\alpha=1}^{n_L-3}(b_{\alpha}|\mu_{\alpha})\prod_{i=1}^{n_L}\cV^{a}_{i}\; \prod_{\beta=1}^{n_R-3}(b_{\beta}|\mu_{\beta})\prod_{j=1}^{n_R}\cV^{a}_{j} \; (b_{a}|\mu_{a})\;\cP_{a}\cV^{f}_{a}\;f(q)\d q\;(b_{b}|\mu_{b})\;\cP_{b}\cV^{g}_{b},
\ee
where the puncture operators enforce the double trace condition.  Under a re-scaling of $\mathsf{x}_{a}\rightarrow \lambda\mathsf{x}_{a}$, it follows that
\begin{equation*}
(b_{a}|\mu_{a})\;\cP_{a}\cV^{f}_{a}\xrightarrow{\mathsf{x}_{a}\rightarrow\lambda\mathsf{x}_{a}} \lambda\;(b_{a}|\mu_{a})\;\cP_{a}\cV^{f}_{a},
\end{equation*}
since the combination $\cP_{a}\cV^{f}_{a}$ has conformal weight $-1$.  Likewise, under the re-scaling $\mathsf{y}_{b}\rightarrow\tilde{\lambda}\mathsf{y}_{b}$ we pick up one power of $\tilde{\lambda}$ from the insertion at $b\in\Sigma_R$.  By \eqref{WF1}, we know that $q$ scales as $q\rightarrow\lambda\tilde{\lambda}q$ so homogeneity of the measure requires that $f(q)=q^{-2}$.  Therefore, a double pole arises in this doubled factorization channel, leaving us with
\begin{multline}\label{CGF3}
\tr(\mathsf{T}^{a_1}\cdots\mathsf{T}^{a_{n_L}})\; \tr(\mathsf{T}^{b_1}\cdots\mathsf{T}^{b_{n_R}}) \prod_{\alpha=1}^{n_L-3}(b_{\alpha}|\mu_{\alpha})\left\la\prod_{i=1}^{n_L}\widetilde{\cV}_{a_i}\right\ra_{S_C} \prod_{\beta=1}^{n_R-3}(b_{\beta}|\mu_{\beta})\left\la\prod_{j=1}^{n_R}\widetilde{\cV}_{b_j}\right\ra_{S_C} \\
\times (b_{a}|\mu_{a})\;\cV^{f}_{a}\;\frac{\d q}{q^2}\;(b_{b}|\mu_{b})\;\cV^{g}_{b},
\end{multline}
for \eqref{CGF2} at the boundary divisor $\mathfrak{D}^{\mathrm{sep}}$.  

Hence, multi-trace terms in the Berkovits-Witten model will always have a factorization channel containing a double pole in $q$.  This channel is the one which is compatible with the color trace, leading to a doubled factorization of both the worldsheet current algebra and the scattering amplitude as a whole.  In the field theory context, this double pole is interpreted as a propagator of the form $p^{-4}$ (see Appendix \ref{QProps}).  This is consistent with a fourth-order theory, so we confirm the claim that $\cV^{f}$, $\cV^{g}$ correspond to conformal gravity degrees of freedom \cite{Berkovits:2004jj}.  Clearly, no such channels exist for the single trace terms which define the planar limit, since every boundary divisor cuts the trace. 

This raises many questions upon comparison with conventional string theory, where higher-order poles have a very different interpretation and factorization of gauge theory amplitudes does not produce non-unitary behavior.  We address these questions below in Section \ref{StrComp}.  For the sake of concreteness, we prove that the multi-trace terms in the Berkovits-Witten twistor-string factorize like $q^{-2}$ at the level of the amplitude itself in Appendix \ref{DTF}.  While the whole point of worldsheet factorization at genus zero is to avoid such computations, it may be useful for some readers to see the double pole emerging explicitly as a result of the structure of the worldsheet current correlator.


\subsection{Skinner's theory and $\cN=8$ supergravity}

Now we turn to the Skinner twistor-string at genus zero.  Although adding worldsheet gravity to this theory by hand results in a non-vanishing central charge, this is only at the expense of an overall factor at genus zero.  Recall that scattering amplitudes for a rational worldsheet are computed by 
\be{Sgz1} 
\cM^{(0)}_{n,d}=\int\frac{\prod_{a=0}^{d}\d^{4|8}U_{a}}{\mathrm{vol}\;\C^{*}}\left\la\prod_{\alpha=1}^{n-3}(b_{\alpha}|\mu_{\alpha})          
\prod_{i=1}^{n}\cV^{h}_i \prod_{j=1}^{n-d-2}\widetilde{\Upsilon}_{j} \prod_{k=1}^{d}\Upsilon_{k}\right\ra.
\ee
As in our discussion of the Berkovits-Witten twistor-string, we are interested in studying the behavior of $\cM^{(0)}_{n,d}$ near the separating boundary divisor $\mathfrak{D}^{\mathrm{sep}}$.  Recall that the full worldsheet of Skinner's theory is the $(1|2)$-dimensional supermanifold $\mathrm{X}$; in the factorization limit, we only need to consider the bosonic body $\Sigma$, though.  This is due to the fact that $\mathrm{X}$ is \emph{split}: all the fermionic degrees of freedom live in the integrable sheaf $\mathcal{D}$ which plays a trivial role in the degeneration of the worldsheet.  This should be contrasted against the situation in the RNS super-string, where the worldsheet is a \emph{super} Riemann surface (which is non-split for generic genus).  In that case, factorization entails looking at a boundary divisor in the moduli space of super Riemann surfaces, and the local model for the worldsheet must take into account the non-trivial supergeometry (c.f., \cite{Witten:2012ga}).

\begin{figure}[t]
\centering
\includegraphics[width=5.5 in, height=1.2 in]{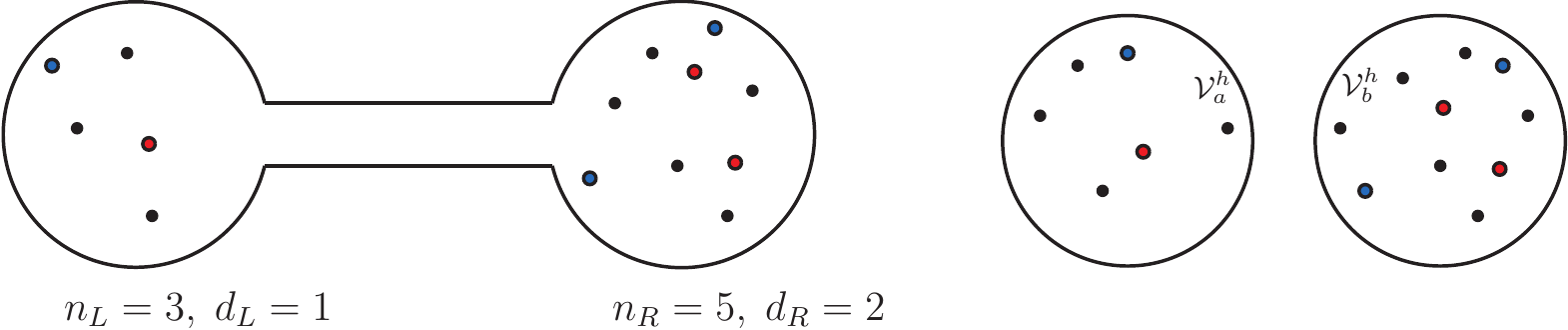}\caption{\small{\textit{The amplitude $\cM_{8,3}$ factorizes into $\cM_{4,1}$ and $\cM_{6,2}$ in Skinner's twistor-string. Black vertices are external vertex operators, red vertices are PCOs $\Upsilon$, and blue vertices are PCOs $\widetilde{\Upsilon}$.}}}\label{TSFac3}
\end{figure}

Once again, we assume that the external vertex operators $\cV^{h}_{i}$ are distributed over the two branches of the degenerate worldsheet as $n_{L}+n_{R}=n$ with $n_{L},n_{R}\geq 2$.  The only new subtlety in Skinner's theory is the presence of the PCOs $\widetilde{\Upsilon}$ and $\Upsilon$ for the $\beta\gamma$- and $\mu\nu$-systems respectively.  Suppose we have $r$ of the $\widetilde{\Upsilon}_{j}$ and $s$ of the $\Upsilon_{k}$ on $\Sigma_{L}$ as we approach the separating divisor.  It is easy to see that the resulting worldsheet correlator on $\Sigma_L$ will vanish unless $r$ and $s$ are appropriately chosen to fix the zero modes of the ghost fields $\gamma$ and $\mu$.  Indeed, to build a top-degree form on the space of fermionic automorphisms and deformations on $\Sigma_L$, we must have $r=n_{L}-d_{L}-1$ and $s=d_{L}$, and likewise on $\Sigma_R$.  So at genus zero, we assume that the PCOs are distributed between $\Sigma_L$ and $\Sigma_R$ in this fashion; an example is illustrated in Figure \ref{TSFac3}.  At higher genus potential subtleties regarding this assumption could arise (e.g., the amplitude is non-zero for an incorrect distribution of the PCOs).  The interested reader may consult section 6.3.6 of \cite{Witten:2012bh} for further discussion.

At this point, the argument for deriving the factorization behavior at genus zero runs as before.  Using the local model \eqref{WF1} for the worldsheet $\Sigma$ as we approach $\mathfrak{D}^{\mathrm{sep}}$, the worldsheet correlator becomes
\begin{multline}\label{Sgz2}
\prod_{\alpha=1}^{n_{L}-3}(b_{\alpha}|\mu_{\alpha})\prod_{i=1}^{n_{L}}\cV_{i}^{h} \prod_{j=1}^{n_{L}-d_{L}-1}\widetilde{\Upsilon}_{j} \prod_{k=1}^{d_{L}}\Upsilon_{k} \; \prod_{\beta=1}^{n_{R}-3}(b_{\beta}|\mu_{\beta}) \prod_{l=1}^{n_{R}}\cV_{l}^{h} \prod_{m=1}^{n_{R}-d_{R}-1}\widetilde{\Upsilon}_{m} \prod_{r=1}^{d_{R}}\Upsilon_{r} \\
\times \sum_{\mathrm{states}}(b_{a}|\mu_{a})\cO_{a}\;f(q)\;\d q\;(b_{b}|\mu_{b})\cO_{b}.
\end{multline}
The sum over states results in a simple pole $f(q)=q^{-1}$ when $\cO_{a}=\cV^{h}_{a}$ and $\cO_{b}=\cV^{h}_{b}$, since these are the only physical vertex operators and have conformal weight zero.  It is also clear that there is no potential for higher-order poles in the modulus $q$, since there is no analogue of multi-trace terms in this theory.

Taking the residue of the simple pole in \eqref{Sgz2} gives the structure of two on-shell worldsheet correlation functions; on $\Sigma_L$ we have
\be{Sgz3}
\left\la\Sigma_{L}\right\ra =\left\la \prod_{\alpha=1}^{n_{L}-2}(b_{\alpha}|\mu_{\alpha})          
\prod_{i=1}^{n_{L}+1}\cV^{h}_i \prod_{j=1}^{n_{L}-d_{L}-1}\widetilde{\Upsilon}_{j} \prod_{k=1}^{d_{L}}\Upsilon_{k}\right\ra,
\ee
where the $\alpha=n_{L}-2$ and $i=n_L+1$ factors correspond to the new insertions at $a\in\Sigma_L$.  The correlation function on $\Sigma_{R}$ is similar in structure.

To deal with the map moduli, we again write the measure near the boundary of the moduli space as
\begin{equation*}
\frac{\prod_{a=0}^{d}\d^{4|8}U_{a}}{\mathrm{vol}\;\C^{*}}=\frac{\d^{4|8}U_{\bullet}}{\mathrm{vol}\;\C^{*}} \prod_{a=1}^{d_{L}}\d^{4|8}U_{a} \prod_{b=1}^{d_{R}}\d^{4|8}W_{b},
\end{equation*}
with the node at $q=0$ mapped to $U_{\bullet}\in\PT$.  We also insert a clever factor of unity, which now takes the form 
\begin{equation*}
1=\int \D^{3|8}W_{\bullet}\;\D^{3|8}Z\;\bar{\delta}^{3|8}_{4,0}(Z,U_{\bullet})\;\bar{\delta}^{3|8}_{0,4}(Z,W_{\bullet}),
\end{equation*}
where the subscripts on $\bar{\delta}^{3|8}$ denote the weights in its two arguments:
\begin{equation*}
 \bar{\delta}^{3|8}_{0,4}(Z,Y)\equiv \int_{\C}\frac{\d s}{s^5}\bar{\delta}^{4|8}(Z+s Y).
\end{equation*}

This gives the desired factorization properties for the original amplitude:
\be{Sgz4}
\cM^{(0)}_{n,d}\xrightarrow{q=0} \int \D^{3|8}Z\;\cM^{(0)}_{n_{L}+1,d_{L}}\;\cM^{(0)}_{n_{R}+1,d_{R}},
\ee
where we again absorb the delta functions $\bar{\delta}^{3|8}_{4,0}(Z,U_{\bullet})$ and $\bar{\delta}^{3|8}_{0,4}(Z,W_{\bullet})$ into the new external states $\cV^{a}_{a}$ and $\cV^{a}_{b}$ respectively.  This constitutes a proof that the Cachazo-Skinner formula has the correct multiparticle factorization properties, and should be contrasted with the proof at the level of the amplitude itself \cite{Cachazo:2012pz}.  In the latter, one introduces a scale into the local model of the worldsheet and then must trace the complicated scale dependence of the various determinants and measures appearing in \eqref{CSi} in order to extract the analogue of a $q^{-1}$ pole.  By working at the level of the worldsheet \emph{before} any Wick contractions or integrals have been performed, we avoid these issues entirely and factorization becomes obvious as a property of the theory.


\subsection{Comparison with string theory}
\label{StrComp}

Having focused so far on worldsheet factorization in twistor-string theory, it is worth taking a moment to contrast this picture with what happens in ordinary string theory.  The factorization properties of bosonic string theory were first studied at the level of the worldsheet long ago (c.f., \cite{Polchinski:1988jq}), and worldsheet factorization extends to the RNS superstring as well (c.f., \cite{Witten:2012bh}).  However, the interpretation of the modulus $q$ differs subtly between twistor-string and conventional string theory.  This is due to the nature of the states which appear in the respective theories.

In this section, we clarify the distinctions between worldsheet factorization of conventional string theory and twistor-strings.  As a concrete example of ordinary string theory, consider a closed bosonic string with matter action\footnote{Formally, the twistor-string is most closely related to a heterotic (or twisted $(0,2)$) string theory \cite{Mason:2007zv, ReidEdwards:2012tq}.  For the purposes of this section, the chiral sector of closed bosonic strings suffices and the salient points are obviously true for open/closed superstrings, the heterotic string, or the pure spinor formalism as well.}
\be{bostring}
S=\frac{-1}{4\pi\alpha'}\int_{\Sigma}\partial X_{\mu}\;\dbar X^{\mu}+\;S_{C}.
\ee
Here $\mu=1,\ldots,D$ is an index for the target space (which we denote by $M$), $X\in\Omega^{0}(\Sigma, T_M)$, and we contract indices using a flat target space metric.  The action $S_C$ for a worldsheet current algebra is chosen in accordance with the various dimensions to cancel anomalies after gauge fixing.

Consider worldsheet factorization at genus zero in this theory with all external states represented by gauge theory vertex operators:
\be{gvert}
\cV=\oint c\;\varepsilon_{\mu}\;\dbar X^{\mu}\wedge j\;\e^{i p\cdot X},
\ee
where $j$ is the current for the worldsheet gauge algebra and we have chosen a chiral representation to parallel the twistor-string.\footnote{One could just as easily take the usual $\cV=c\tilde{c}\varepsilon_{\mu} \dbar X^{\mu}j \e^{ip\cdot X}$, at the expense of introducing the anti-holomorphic $\tilde{c}$ and $\tilde{b}$ at each stage of the discussion.}  This is a homogeneous scalar on the worldsheet, with polarization tensor $\varepsilon_{\mu}$ and momentum $p_{\mu}$ satisfying $p^{2}=0=p\cdot\varepsilon$.  The worldsheet correlation function for a genus zero scattering amplitude is given by
\be{stgz1}
\left\la \prod_{\alpha=1}^{n-3}(b_{\alpha}|\mu_{\alpha})\;\prod_{i=1}^{n}\cV_{i}\right\ra,
\ee
which is integrated over the moduli space $\overline{M}_{0,n}$ of the bosonic string.

The worldsheet factorization of \eqref{stgz1} proceeds in the manner described in Section \ref{FL}.  As we approach the boundary divisor $\mathfrak{D}^{\mathrm{sep}}$, the worldsheet develops a long tube and we get a sum over states inserted at either end of this tube.  Using the local model \eqref{WF1} as we approach the boundary, the correlator takes the form:
\be{strfac1}
\prod_{\alpha=1}^{n_{L}-3}(b_{\alpha}|\mu_{\alpha})\prod_{i=1}^{n_{L}}\cV_{i}\:\prod_{\beta=1}^{n_{R}-3}(b_{\beta}|\mu_{\beta}) \prod_{j=1}^{n_{R}}\cV_{j} \sum_{\mathrm{states}}(b_{a}|\mu_{a})\cO_{a}\;f(q)\;\d q\;(b_{b}|\mu_{b})\cO_{b},
\ee
As in the Berkovits-Witten twistor-string, the contribution of the worldsheet current algebra to this amplitude can be broken up according to the number of traces.  Away from the planar limit, we could worry that multi-trace terms may lead to higher order poles in the modulus $q$ as they did in the twistor-string.

For example, suppose we consider a double-trace contribution and focus on the factorization channel which is compatible with the trace structure.  As we approach the factorization limit, this will include terms where a graviton vertex operator
\be{graviton}
\cV^{\mathrm{grav}}=\oint c\;\varepsilon_{\mu\nu}\;\partial X^{\mu}\;\dbar X^{\nu}\;\e^{i p\cdot X},
\ee
is inserted for $\cO_{a}$, $\cO_b$.  This gives a contribution to the correlator near the boundary:
\begin{multline*}
\tr(\mathsf{T}^{a_1}\cdots\mathsf{T}^{a_{n_L}})\; \tr(\mathsf{T}^{b_1}\cdots\mathsf{T}^{b_{n_R}}) \prod_{\alpha=1}^{n_L-3}(b_{\alpha}|\mu_{\alpha})\prod_{i=1}^{n_L}\widetilde{\cV}_{a_i} \prod_{\beta=1}^{n_R-3}(b_{\beta}|\mu_{\beta})\prod_{j=1}^{n_R}\widetilde{\cV}_{b_j} \\
\times (b_{a}|\mu_{a})\;\cV^{\mathrm{grav}}_{a}\;f(q)\d q\;(b_{b}|\mu_{b})\;\cV^{\mathrm{grav}}_{b},
\end{multline*}
where $\widetilde{\cV}_{a_i}$ is \eqref{gvert} stripped of a generator $\mathsf{T}^{a_i}$.  

At first glance, we may be tempted to say that this is equivalent to the situation in \eqref{CGF2} for the twistor-string, so we should get a double pole in $q$.  But this is simply not true: in the twistor-string the double trace factorized the correlation function of external states because the only Wick contractions were from the worldsheet current algebra.  In string theory, there is a non-trivial OPE between the $X$s:
\begin{equation*}
X^{\mu}(z)\;X^{\nu}(z')\sim -\eta^{\mu\nu}\;\ln|z-z'|^{2}, 
\end{equation*}
so restricting to a double-trace contribution does \emph{not} factorize the correlator.  Each $\cV$ can Wick contract with every other external state away from the boundary divisor.

Hence, the doubled factorization structure we found in the twistor-string does not exist in ordinary string theory because of the basic structure of the vertex operators.  So when factorizing a scattering amplitude in string theory, all terms in \eqref{strfac1} for which $\cO_{a}$ and $\cO_{b}$ are vertex operators of conformal dimension zero will have $f(q)=q^{-1}$ by homogeneity of the moduli space measure.  In the multi-trace example, this is consistent with the vertex operators $\cV^{\mathrm{grav}}$ being gravitons.  

However, the space of bosonic string states also contains the tachyon vertex operator:
\be{tach}
\cV^{\mathrm{tach}}=c\;\e^{ip\cdot X},
\ee
so there is a term contributing to \eqref{strfac1} with $\cO_{a,b}=\cV^{\mathrm{tach}}_{a,b}$.  This operator takes values in $T_{\Sigma}$ (i.e, $\cV^{\mathrm{tach}}$ has conformal weight $-1$), so under a scaling $\mathsf{x}_{a}\rightarrow\lambda\mathsf{x}_{a}$ we have $\cV^{\mathrm{tach}}_{a}\rightarrow\lambda\cV^{\mathrm{tach}}_{a}$.  Homogeneity of the measure requires that $f(q)=q^{-2}$ when tachyon operators are inserted, and we get a contribution near the boundary divisor of the form:
\begin{equation*}
 \prod_{\alpha=1}^{n_{L}-3}(b_{\alpha}|\mu_{\alpha})\prod_{i=1}^{n_{L}}\cV_{i}\:\prod_{\beta=1}^{n_{R}-3}(b_{\beta}|\mu_{\beta}) \prod_{j=1}^{n_{R}}\cV_{j}\; (b_{a}|\mu_{a})\cV^{\mathrm{tach}}_{a}\;\frac{\d q}{q^{2}}\;(b_{b}|\mu_{b})\cV^{\mathrm{tach}}_{b}
\end{equation*}
We saw in an earlier example that inserting identity operators led to such a double pole; another way of saying this is that the identity operator is the tachyon operator at zero momentum.\footnote{Although our focus has been at genus zero, the tachyon enters in this way for the factorization of string theory amplitudes at higher genus as well \cite{D'Hoker:2001qp, D'Hoker:2002gw}.}

So in ordinary string theory, finding a double pole in the factorization limit of a scattering amplitude signals the presence of a tachyon.  An equivalent way of seeing this is to write the measure on the moduli space near the factorization divisor $\mathfrak{D}^{\mathrm{sep}}$ in terms of the string propagator (c.f., \cite{Green:1987} Chapter 7 and \cite{Polchinski:1998} Chapter 9), which in our language takes the form:
\be{stprop}
\Delta = b_{0}\;\int_{0}^{1}\frac{\d q}{q}\;q^{L_{0}}=\frac{b_{0}}{L_{0}},
\ee
where $b_{0}$ is a zero-mode of the $b$-antighost, the limits on the integral correspond to integrating over the moduli $q$ (only the $q\sim 0$ region is relevant for the IR behavior we are interested in), and $L_{0}$ is the Hamiltonian which acts as the conformal weight operator.  The form of \eqref{stprop} illustrates that a simple pole in the modulus $q$ is equivalent to $L_{0}\cV=0$; for on-shell states this means that the momentum being carried by the operator in the factorization limit is null.  So a simple pole in $q$ is the same as a $p^{-2}$ propagator vanishing in field theory (see section 6 of \cite{Witten:2012bh} for further discussion).

However, a double pole in string theory means that $L_{0}\cV=-1$ (i.e., we have a tachyon) rather than a higher-order propagator (e.g., $p^{-4}$) going on-shell as we claimed for the twistor-string.  Why the apparent contradiction?  First, note that $L_{0}\cV=0$ for \emph{all} BRST-closed states in twistor-string theory: there is no analogue of the tachyon operator, and states are specified by a generic cohomology class.  Furthermore, when momentum eigenstates are chosen for the external states, all the bosonic moduli integrals are fixed by delta-functions \cite{Roiban:2004yf}.  This means that there is a direct relationship between $q$ and the external momenta; in the case of multiparticle factorization, one finds that
\begin{equation*}
 P_{L}^{2}=\left(\sum_{i=1}^{n_{L}}p_{i}\right)^{2}\propto q,
\end{equation*}
for both the Berkovits-Witten \cite{Vergu:2006np} and Skinner twistor-strings \cite{Cachazo:2012pz}.  We review this in Appendix \ref{QProps}.  

So in twistor-string theory, there is a direct relationship between inverse powers of $q$ and propagators of the form $p^{-2}$ as we approach the factorization limit.  In ordinary string theory, on the other hand, this relationship only holds for vertex operators satisfying the massless condition $L_0\cV=0$; this is manifest in the functional form of the string propagator \eqref{stprop}.  Higher-order poles are indicative of tachyonic vertex operators, which are removed by the GSO projection \cite{Gliozzi:1976jf} in passing to superstring theory.


\section{Discussion \& Conclusion}
\label{Concl}

In this paper, we investigated the factorization properties of scattering amplitudes in twistor-string theories at the level of the worldsheet.  The power of this approach is in manifesting the correct singularity structure of the amplitudes at the boundary of the moduli space, without having to first compute the full amplitude itself.  In essence, this provides a simple and manifestly geometric proof of multiparticle factorization for the amplitudes.  Combined with the correct three-particle amplitudes, worldsheet factorization proves that the genus zero scattering amplitudes of the theories are correct by BCFW recursion.  Hence, the worldsheet approach sidesteps the complicated proofs of factorization required at the level of the scattering amplitude formulae themselves (c.f., \cite{Skinner:2010cz, Cachazo:2012pz}).

Furthermore, we also saw how to understand the appearance of higher-order factorization poles associated with conformal gravity degrees of freedom in the Berkovits-Witten twistor-string.  These arose by considering factorization channels which were compatible with the color structure of multiple trace contributions to the scattering amplitude.  Since the worldsheet correlator is doubly factorized in this limit (once at the level of the color structure and again at the level of the worldsheet), we found a double pole indicative of a fourth-order theory.  We have also derived this double pole structure at the level of the amplitude itself in Appendix \ref{DTF}, and saw why the same structure does not arise in conventional string theory.   

Of course, there are many future directions and questions raised by this work.  Therefore, let us conclude with a brief discussion of some of these issues and their prospects for future development.

\subsection*{Beyond genus zero}

Perhaps the most obvious question to ask is: what happens for higher-genus worldsheets?  Even without delving into precise calculations, there are a few immediate observations that we can make based on our studies at genus zero.  In the Berkovits-Witten theory, the planar limit for gauge theory external states is no longer sufficient to suppress conformal gravity degrees of freedom, since we have no control over the states running in the `loops' of a higher genus Riemann surface.  Hence, we know that the double poles associated with conformal supergravity will appear in the factorization limit regardless of the rank of the gauge group.  This is just another way of stating the basic problem with the original twistor-string: conformal gravity degrees of freedom contaminate the gauge theory scattering amplitudes beyond tree-level.

The situation in Skinner's theory is more complicated.  Recall that we rather brutally added worldsheet gravity into the theory by hand in order to study worldsheet factorization.  The central charge of the theory becomes non-vanishing, and the integral over non-zero-modes of the $bc$-ghost system produces a factor of $\mathrm{det}'\dbar_{T_\Sigma}$, which is a section of a determinant line bundle over $\overline{M}_{g}$.  When $g=0$, this is just an overall constant, but for $g>0$ the topology of this line bundle will become non-trivial.  This means that $\mathrm{det}'\dbar_{T_\Sigma}$ will have some dependence on the moduli of $\Sigma$, and in particular may depend on the modulus $q$ which is transverse to the factorization boundary divisor.  Novel $q$-dependence will clearly alter the pole structure we obtain in the measure for the moduli space near the divisor, and in turn the factorization behavior of the amplitude.

Hence, it seems that our prescription for adding worldsheet gravity to Skinner's theory by hand will break down without doing explicit calculations.  This can be seen as a potential problem even at the level of writing down a formula for the genus one amplitude: how do we specify the integral over the modular parameter $\tau$ in the absence of worldsheet gravity?  Since the theory includes a charged, rank-two $\beta\gamma$-system, it does include a prescription for integrating over the moduli of $\SL(2,\C)$-bundles on $\Sigma$ (c.f., section 5 of \cite{Skinner:2013xp}).  This could lead to a different divergence structure (in both the UV and IR) than conventional string theory.  The worldsheet factorization perspective may prove a useful tool for investigating these structures in the future.

As noted in Section \ref{WSG}, one alternative for building a measure on the worldsheet moduli space at higher genus is the introduction of an effective $b$-ghost, akin to the pure spinor formalism for string perturbation theory \cite{Berkovits:2000fe}.  This would be a composite field $b^{\mathrm{eff}}$ which is fermionic, takes values in $K^{2}_{\Sigma}$, and obeys $\{Q,b^{\mathrm{eff}}\}=T$, where $T$ is the stress-energy tensor of the twistor-string:
\begin{equation*}
T=\la Y,\partial Z\ra +\frac{i}{2}\la\rho^{a},\partial\rho_{a}\ra+\frac{i}{2}\left(\gamma^{a}\partial\beta_{a}+3\beta_{a}\partial\gamma^{a}\right)
+\frac{i}{2}\left(\mathrm{m}_{ab}\partial\mathrm{n}^{ab}+\mathrm{m}\partial\mathrm{n}\right)+\frac{i}{2}\mu_{a}\partial\nu^{a}.
\end{equation*}
Unfortunately, there is currently no known way to construct $b^{\mathrm{eff}}$ for Skinner's twistor-string; so we have no consistent mechanism for including higher-genus moduli integrals other than by hand.  It is worth noting that if such a prescription were to exist, it would essentially promote Skinner's theory to a proper string theory--in which case it would be free of UV divergences.  Combined with invariance under the Teichm\"uller modular group and appropriate factorization properties, this would seem to indicate that $\cN=8$ supergravity is UV finite.  While the theory certainly possesses surprisingly good UV behavior (c.f., \cite{Bern:2011qn}), many arguments based on the presence of symmetry-preserving counter-terms indicate that divergences should appear in the neighborhood of seven loops \cite{Howe:1980th, Bjornsson:2010wm, Bossard:2011tq, Beisert:2010jx, Kallosh:2011dp}.  One may speculatively suggest that the inability to consistently incorporate worldsheet gravity in Skinner's theory is due to the eventual appearance of UV divergences in supergravity.     

Finally, at genus one worldsheet factorization must account not only for the separating factorization divisor we considered in this paper, but also the \emph{non-separating} degeneration which reduces the genus by one.  This means that there is a novel geometric feature of worldsheet factorization that does not appear at genus zero; even in conventional string theory this has received only limited treatment (c.f., \cite{D'Hoker:1994yr, D'Hoker:2002gw}).  Treating these non-separating degenerations formally will require a careful understanding of the role of worldsheet gravity and the spectrum of string states on higher-genus worldsheets.   

\subsection*{Other formulae}

The recent flurry of progress in our understanding of scattering amplitudes in gauge theory and gravity has reached its latest incarnation in the remarkable formulae of Cachazo, He, and Yuan (CHY) for the tree-level scattering of pure Yang-Mills theory and gravity in arbitrary dimension \cite{Cachazo:2013gna, Cachazo:2013hca}.  Like the RSV or Cachazo-Skinner formulae, the CHY expressions are remarkably compact--indicating structures in play which are much simpler than traditional Feynman diagram techniques for a wide variety of massless theories in any dimension.  Furthermore, the CHY formulae encode deep information about the color kinematics correspondence and even provide a novel proof of the BCJ relations \cite{Cachazo:2013iea}.

From the perspective of this paper, one of the most striking facts about the CHY formulae is that they express scattering amplitudes as an integral over the moduli space of rational curves with marked points.  For instance, the CHY formula for the tree-level S-matrix of pure Yang-Mills theory in the planar limit reads \cite{Cachazo:2013hca}:
\be{CHYi}
\cA_{n}=\int \frac{\prod_{i=1}^{n}\D\sigma_i}{\mathrm{vol}\;\SL(2,\C)}|\sigma_i\sigma_j\sigma_k|\prod_{m\neq i,j,k}\delta(\mathcal{S}_m)\frac{\tr(\mathsf{T}^{a_1}\mathsf{T}^{a_2}\cdots\mathsf{T}^{a_n})}{(\sigma_1\sigma_2)(\sigma_2\sigma_3)\cdots(\sigma_n\sigma_1)}\mathrm{Pf}'\Psi.
\ee
The product of delta functions enforces $n-3$ of the `scattering equations' \cite{Cachazo:2013gna}:
\begin{equation*}
\mathcal{S}_i\equiv\sum_{j\neq i}\frac{p_i\cdot p_j}{(\sigma_i\sigma_j)}=0.
\end{equation*}
The trace is over the generators of the gauge group, and $\Psi$ is a $2n\times 2n$ skew-symmetric matrix depending on the $\sigma_i$ as well as the $p_i$ and polarization vectors $\epsilon_i$.  The reduced Pfaffian appearing in \eqref{CHYi} is defined as
\begin{equation*}
\mathrm{Pf}'\Psi\equiv (-1)^{i+j}\frac{\mathrm{Pf}\Psi^{ij}_{ij}}{(\sigma_i\sigma_j)}.
\end{equation*}
It can be shown that this formula is independent of the choice of $\sigma_{i},\sigma_{j},\sigma_{k}$ as well as the two rows and columns removed from $\Psi$.

As with the RSV and Cachazo-Skinner formulae, the veracity of \eqref{CHYi} is established by studying its multiparticle factorization, collinear, and soft limits.  Once again, factorization is the most non-trivial and difficult of these properties to check.  Clearly, a worldsheet perspective would serve the desirable purpose of simplifying multiparticle factorization as well as raising the possibility of extending formulae such as \eqref{CHYi} beyond tree-level.

There are many indications that such a worldsheet theory must exist.  Not least among them is the fact that these formulae have the structure of a worldsheet correlator built in: the integral over $\overline{M}_{0,n}$ and the quotient by the volume of $\SL(2,\C)$ are indicative of closed worldsheet gravity.  There are also footprints of worldsheet supersymmetry: the rows and columns removed from the matrix $\Psi$ are indicative of a mix of integrated and fixed vertex operators, just like rows and columns removed from $\HH$ and $\HH^{\vee}$ in the Cachazo-Skinner formula, and it is not hard to reverse engineer what the manifestly permutation symmetric form of the CHY formulae should be.  It would be intriguing to know what role--if any--worldsheet factorization has to play in studying the structures underlying these formulae.

\subsection*{Disconnected Prescription}

An important tool to emerge from recent advances in our understanding of scattering amplitudes in Yang-Mills theory is the MHV (or CSW) formalism \cite{Cachazo:2004kj}.  This is a set of simple Feynman rules for gauge theory, whose vertices are MHV amplitudes (extended off-shell) tied together by $p^{-2}$ propagators.  The MHV formalism provides a method for computing tree-level amplitudes that is much more efficient than traditional Lagrangian-based Feynman rules, and can also be extended effectively to loop-level computations \cite{Brandhuber:2004yw, Lipstein:2012vs, Lipstein:2013xra}.  It is therefore natural to ask whether a MHV formalism exists for gravity; unfortunately, it is known that the na\"ive extension of the off-shell prescription for the vertices from Yang-Mills theory fails when applied to gravity \cite{BjerrumBohr:2005jr, Bianchi:2008pu}.

Nevertheless, there are strong indications that a MHV formalism for gravity does indeed exist (c.f., \cite{Adamo:2013tja, Penante:2012wd}), and the worldsheet factorization of Skinner's twistor-string at tree-level adds to this growing evidence.  Indeed, the intuition for the MHV formalism in gauge theory first came from interpreting Witten's twistor-string at the boundary of the moduli space.  This is known as the `disconnected prescription': instead of considering the scattering amplitude for degree $d$ maps, one considers the degeneration of the correlator on $d$ degree-one components which are linked by $p^{-2}$ propagators \cite{Gukov:2004ei}.  Repeatedly applying genus zero factorization to Skinner's theory reveals the same structure as an easy consequence of worldsheet factorization.

\begin{figure}[t]
\centering
\includegraphics[width=3.5 in, height=1.5 in]{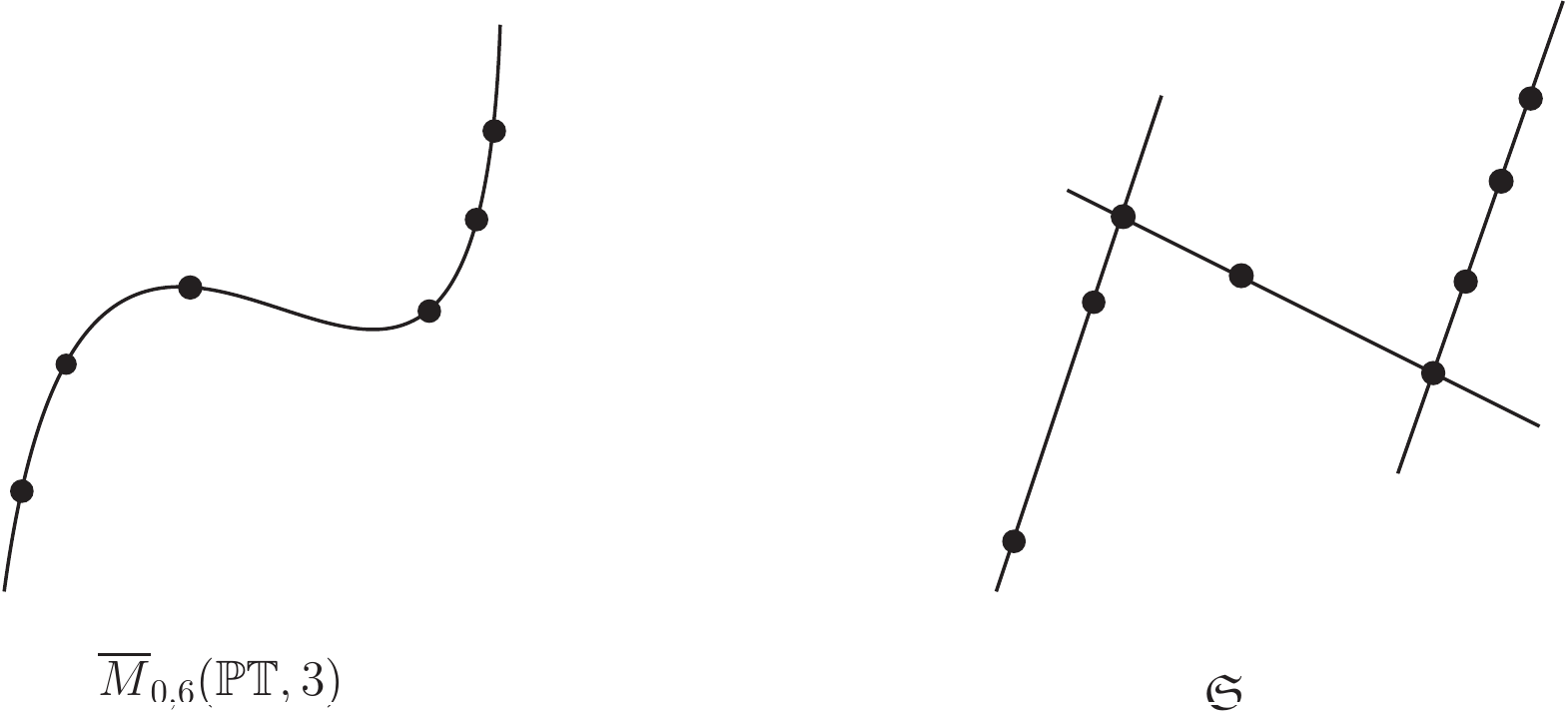}\caption{\small{\textit{The degeneration of a $d=3$ map into three intersecting $d=1$ components.}}}\label{Disconn}
\end{figure}

To be more precise, consider a $n$-point scattering amplitude in Skinner's theory for maps of degree $d>1$.  We want to consider a co-dimension $d-1$ subvariety of the moduli space $\overline{M}_{0,n}(\PT,d)$ given by
\begin{equation*}
 \mathfrak{S}\cong\overline{M}_{0,n_1}(\PT,1)\times\overline{M}_{0,n_2}(\PT,1)\times\cdots\times\overline{M}_{0,n_{d}}(\PT,1).
\end{equation*}
In the language of scattering amplitudes, this corresponds to degenerating a N$^{d-1}$MHV amplitude into $d$ MHV amplitudes, in accordance with the general prescription of the MHV formalism.  An example for the six-point N$^2$MHV amplitude is shown in Figure \ref{Disconn}.  We want to know what the structure of the worldsheet correlator is as we approach $\mathfrak{S}$ in the moduli space.

The answer is provided by our investigation of worldsheet factorization in this paper.  In particular, we can reach $\mathfrak{S}$ by first approaching the separating divisor
\begin{equation*}
 \mathfrak{D}^{\mathrm{sep}}_{1}\cong\overline{M}_{0,n-n_1+2}(\PT,d-1)\times\overline{M}_{0,n_1}(\PT,1).
\end{equation*}
We have shown that as we approach $\mathfrak{D}^{\mathrm{sep}}_{1}$, the measure on the moduli space develops a simple pole of the form $q_{1}^{-1}$, where $q_{1}$ is a local coordinate on $\overline{M}_{0,n}(\PT,d)$ transverse to $\mathfrak{D}^{\mathrm{sep}}_{1}$.  In momentum space, this corresponds to a $p^{-2}$ propagator.  Now, staying near $\mathfrak{D}^{\mathrm{sep}}_{1}$, we move to another separating divisor--now inside $\overline{M}_{0,n-n_1+2}(\PT,d-1)$:
\begin{equation*}
\mathfrak{D}^{\mathrm{sep}}_{2}\cong\overline{M}_{0,n-n_1-n_2+4}(\PT,d-2)\times\overline{M}_{0,n_2}(\PT,1).
\end{equation*}
Once again, worldsheet factorization tells us that the measure on the moduli space near $\mathfrak{D}^{\mathrm{sep}}_{2}$ has a simple pole of the form $q_2^{-1}$, again corresponding to a $p^{-2}$ propagator.

Clearly, we can proceed inductively to obtain $\mathfrak{S}$ with a measure that has $d-1$ poles of the form $q^{-1}$.  If we denote the scattering amplitude in shorthand by a top-degree form $\omega_{n,d}$ to be integrated over the moduli space,
\begin{equation*}
 \cM_{n,d}=\int\limits_{\overline{M}_{0,n}(\PT,d)}\omega_{n,d},
\end{equation*}
then as we approach the subvariety $\mathfrak{S}$ this measure looks like
\begin{equation*}
 \left.\omega_{n,d}\right|_{\mathfrak{S}}=\omega_{n_1,1}\wedge\frac{\d q_1}{q_1}\wedge\omega_{n_2,1}\wedge\frac{\d q_2}{q_2}\wedge\cdots\wedge\frac{\d q_{d-1}}{q_{d-1}}\wedge\omega_{n_{d},1}.
\end{equation*}
Hence, the residue of $\cM_{n,d}$ evaluated on $\mathfrak{S}$ is just a product of $d$ MHV amplitudes.  But the arguments of \cite{Gukov:2004ei} now imply that this should also be the residue of the amplitude computed in the MHV formalism.

While it certainly doesn't provide an explicit prescription for the MHV formalism itself, this shows that worldsheet factorization is consistent with the structure of a theory for which such a formalism exists.  It is a clear goal of future research to explicitly derive the MHV rules of Einstein gravity on momentum space, perhaps by translating the twistorial rules of \cite{Adamo:2013tja}.

\subsection*{Leading Singularities}

In perturbative quantum field theory, the \emph{leading singularity} of a scattering amplitude is the singularity resulting from putting all internal propagators in a loop amplitude on-shell, giving to a product of on-shell tree-amplitudes as residues \cite{Britto:2004nc}.  For string-theoretic scattering amplitudes, the analogue of a leading singularity is a maximal degeneration of the genus $g$ worldsheet into Riemann spheres (which can themselves be degenerated into 3-point amplitudes).  Worldsheet factorization immediately tells us that this process is the analogue of the field theory leading singularity, since these degenerations correspond to simple poles in the moduli whose singularities are mapped to on-shell propagators in the field theory.  

It has been shown that a twistor-string for $\cN=4$ super-Yang-Mills (without conformal gravity) would be highly constrained--indeed, perhaps determined by--the requirement that its genus $g$ leading singularities mapped onto the $g$-loop leading singularities of the gauge theory \cite{Bullimore:2009cb}.  Combined with the embedding of the twistor-string moduli space into the Grassmannian, this produced sharp constraints on the types of leading singularities which could appear in the gauge theory at higher loop order.  

One may wonder what information about the leading singularities of $\cN=8$ supergravity is carried by Skinner's twistor-string.  On one hand, the difficulties at higher genus discussed above appear to subvert any attempt to extract factorization information at higher genus in the usual way.  Suppose, however, that we are able quotient out by any additional moduli dependence introduced by the conformal anomaly associated with worldsheet gravity.  Assuming that such a construction exists and is consistent, it is not hard to see that extracting the leading singularities from higher genus amplitudes will result in the correct leading singularities of $\cN=8$ supergravity.

So it seems as if Skinner's twistor-string encodes higher-loop information in the guise of leading singularities; the obstruction to accessing this information is tied up in the need to find a consistent prescription for the twistor-string at higher genus.  Hopefully future work on the twistor-string beyond genus zero will shed light on how this can be done.

\acknowledgments

It is my pleasure to thank Eduardo Casali, Yvonne Geyer, Michael Green, Ron Reid-Edwards, and David Skinner for many useful conversations and comments.  I would also like to thank the Simons Center for Geometry and Physics for hospitality while portions of this project were being completed.  This work is supported by a Title A research fellowship at St John's College, Cambridge.


\appendix

\section{Genus Zero Scattering Amplitudes}
\label{GZFs}

In this appendix we compute the genus zero scattering amplitudes for the manifestly permutation symmetric formalism of Skinner's theory, and show that they are equivalent to the Cachazo-Skinner formula.  In particular, the manifestly permutation symmetric setup simply provides a different representation of the same answer.

To begin, we need to compute the genus zero worldsheet correlation function:
\begin{equation*}
 \left\la \prod_{i=1}^{n}V^{h}(\sigma_i)\;\prod_{j=1}^{n-d-2}\widetilde{\Upsilon}(x_j)\;\prod_{k=1}^{d}\Upsilon(y_k)\right\ra.
\end{equation*}
Let's start by considering the contribution from the $\beta\gamma$-system.  The only appearance of $\beta_{a}$ or $\gamma^{b}$ in the correlator is through the delta-function prefactors of the PCOs $\widetilde{\Upsilon}_j$ from \eqref{bPCO} and the $V^{h}_i$ from \eqref{VO}, respectively:
\be{bgcorr1}
\left\la \prod_{i=1}^{n}\delta^{2}(\gamma_i) \prod_{j=1}^{n-d-2}\delta^{2}(\beta_j)\right\ra_{\beta\gamma}.
\ee
Each component of $\gamma^{a}$ has $d+2$ zero-modes at genus zero while $\beta$ has no zero-modes, so all the $\delta^{2}(\beta)$ insertions must be eaten by Wick contractions, and the remaining $d+2$ $\delta^{2}(\gamma)$s are treated by a path integral over zero-modes weighted by the action
\begin{equation*}
I_{\beta\gamma}=\frac{1}{2\pi}\int_{\Sigma}\beta_{a}\dbar\gamma^{a}.
\end{equation*}
In other words, we need to calculate:
\be{bgcorr2}
\int [\D\gamma][\D\beta]\prod_{i=1}^{n}\delta^{2}(\gamma_i)\prod_{j=1}^{n-d-2}\delta^{2}(\beta_j) \e^{-I_{\beta\gamma}}.
\ee

To evaluate \eqref{bgcorr2}, first consider similar correlators in an arbitrary $\beta\gamma$-system having only a single component for each ghost.  If neither $\beta$ nor $\gamma$ have any zero-modes, then 
\begin{equation*}
\la \delta(\gamma_i)\delta(\beta_j)\ra=\int [\D\beta][\D\gamma]\;\e^{-I_{\beta\gamma}}\delta(\gamma_i)\delta(\beta_j)\frac{1}{\mathrm{det}'\dbar}\frac{1}{S(\sigma_{i},x_{j})},
\end{equation*}
where $S(\sigma,\sigma')$ is the propagator of the $\beta\gamma$-system (c.f., \cite{Witten:2012bh} Chapter 10 for more details).  Adding an additional component to each ghost will simply result in a squared answer: $S(\sigma_i,x_j)^{-2}$.

Now, in \eqref{bgcorr2}, each component of $\gamma^{a}$ has $d+2$ zero-modes.  A basis for these zero modes is given by:
\be{gbasis}
\mathcal{Y}_{j}(\sigma)=\frac{\sigma^{\alpha_1}\cdots\sigma^{\alpha_{d+1}}}{\sqrt{\D\sigma}}\in\Gamma(K^{-1/2}_{\Sigma}\otimes\cO(d)),
\ee
where $j=1,\ldots, d+2$ runs over the different choices for the indices $\alpha=0,1$ in the numerator.  This means that we can expand
\begin{equation*}
\gamma_{a}(\sigma_{i})=\sum_{j=1}^{d+2}\Gamma_{aj}\mathcal{Y}_{j}(\sigma_i)+\mbox{non-zero-modes},
\end{equation*}
and \eqref{bgcorr2} gives a Slater determinant
\begin{multline}\label{bgcorr3}
\left\la \prod_{i=1}^{n}\delta^{2}(\gamma(\sigma_i)) \prod_{j=1}^{n-d-2}\delta^{2}(\beta(x_j))\right\ra_{\beta\gamma}=  \\
=\int [\D\beta]_{\mathrm{n.z.m.}}[\D\gamma]_{\mathrm{n.z.m.}}\d^{2}\Gamma_{1}\cdots\d^{2}\Gamma_{d+2}\;\e^{-I_{\beta\gamma}}\prod_{i=1}^{n}\delta^{2}(\gamma_i) \prod_{j=1}^{n-d-2}\delta^{2}(\beta_j) =\frac{1}{|\mathsf{N}|^2},
\end{multline}
where the matrix $\mathsf{N}$ is given by:
\be{Nmat*}
\mathsf{N}=\left( 
\begin{array}{cccccc}
\mathcal{Y}_{1}(\sigma_1) & \cdots & \mathcal{Y}_{d+2}(\sigma_1) & S(\sigma_{1},x_{1}) & \cdots & S(\sigma_{1},x_{n-d-2}) \\
\vdots & & \vdots & \vdots & & \vdots \\
\mathcal{Y}_{1}(\sigma_{n}) & \cdots & \mathcal{Y}_{d+2}(\sigma_{n}) & S(\sigma_{n},x_{1}) & \cdots & S(\sigma_{n}, x_{n-d-2}) 
\end{array}\right).
\ee 

Note that the propagator $S$ is not simply the usual $(\sigma x)^{-1}$, since $\gamma^{a}$ and $\beta_{a}$ are charged under the line bundle $\cL\cong\cO(d)$.  In particular, $\gamma^{a}$ takes values in $K^{-\frac{1}{2}}_{\Sigma}\otimes\cO(d)$ and $\beta_{a}$ takes values in $K^{\frac{3}{2}}_{\Sigma}\otimes\cO(-d)$, so there is an ambiguity in defining the propagator.  This propagator corresponds to $\dbar^{-1}$ acting on forms of weight $d$ taking values in the spin bundle, so there is a $(d+2)$-fold ambiguity which can be fixed by requiring that the propagator vanishes at $d+2$ points on $\Sigma\cong\P^1$:
\be{bgprop}
\left\la\gamma(\sigma)\beta(x)\right\ra=S(\sigma, x)=\frac{(\D x)^{3/2}}{\sqrt{\D\sigma}}\frac{1}{(x\sigma)}\prod_{s=1}^{d+2}\frac{(b_{s}\sigma)}{(b_{s}x)},
\ee
where the $d+2$ reference points $\{b_s\}$ are arbitrarily chosen on $\P^1$, and the worldsheet correlator is independent of the choice of their locations

A similar story holds for the contribution to the genus zero correlator from the $\mu\nu$-system. In this case, we know that each component of $\mu_{a}$ has $d$ zero modes, which are fully saturated by the $\Upsilon$ insertions.  Hence, expanding in a basis of zero modes gives us a simplified Slater determinant:
\be{mncorr1}
\left\la \prod_{k=1}^{d}\delta^{2}(\mu_k)\right\ra_{\mu\nu}=\frac{1}{|y_{1}\cdots y_{d}|^{2}\;\prod_{k=1}^{d}\D y_{k}},
\ee
which is just a Vandermonde determinant in the PCO insertion points.

This leaves us with a reduced version of the initial correlator:
\be{gzz1}
\frac{\prod_{k=1}^{d}\D y_{k}^{-1}}{|\mathsf{N}|^2}\frac{1}{|y_{1}\cdots y_{d}|^{2}} \left\la\prod_{i=1}^{n}h_{i} \prod_{j=1}^{n-d-2} Y_{j\;I}\rho^{I}[Y_{j},\tilde{\rho}]\prod_{k=1}^{d}\la\rho, Z_{k}\ra \tilde{\rho}_{I} Z^{I}_{k}\right\ra,
\ee
where all remaining Wick contractions need to be evaluated in the $\rho\tilde{\rho}$ and $YZ$-systems on the worldsheet.  The index structure of these fields further simplifies the remaining contractions we need to compute.  For instance, suppose Wick contractions of the $\rho\tilde{\rho}$-system were to intertwine insertions from the two different types of PCO.  This would lead to a contribution to \eqref{gzz1} of the form
\begin{equation*}
 \left\la [Y(x_i),\tilde{\rho}]\;\la\rho,Z(y_j)\ra\right\ra_{\rho\tilde{\rho}}.
\end{equation*}

But $[Y,\tilde{\rho}]=Y_{\dot{\alpha}}\tilde{\rho}^{\dot{\alpha}}$ and $\la\rho,Z\ra=\rho_{\alpha}Z^{\alpha}$, while the propagator for the $\rho\tilde{\rho}$-system is
\be{rprop}
\left\la\rho^{I}(x)\;\tilde{\rho}_{J}(y)\right\ra=\delta^{I}_{J}\frac{\sqrt{\D x}\;\sqrt{\D y}}{(x\;y)}.
\ee
This ensures that all such contractions vanish.  A similar argument excludes mixed contractions from the $YZ$-system as well.

With this in mind, the worldsheet correlator \eqref{gzz1} becomes
\be{gzz2}
\frac{\prod_{k=1}^{d}\D y_{k}^{-1}}{|\mathsf{N}|^2}\frac{1}{|y_{1}\cdots y_{d}|^{2}}\left\la\prod_{i=1}^{n}h_{i} \prod_{j=1}^{n-d-2} Y_{j\;I}\rho^{I}[Y_{j},\tilde{\rho}]\right\ra \left\la\prod_{k=1}^{d}\la\rho, Z_{k}\ra \tilde{\rho}_{I} Z^{I}_{k}\right\ra.
\ee
The final factor is exactly the same as the contribution from the $\Upsilon$ PCOs in Skinner's original formulation \cite{Skinner:2013xp}:
\begin{equation*}
\left\la\prod_{k=1}^{d}\la\rho, Z_{k}\ra \tilde{\rho}_{I} Z^{I}_{k}\right\ra=\left|\HH^{\vee}\right|\;\prod_{k=1}^{d}\D y_{k},
\end{equation*}
where $\HH^{\vee}$ is a $d\times d$-matrix with entries
\be{DHM*}
\HH^{\vee}_{kl}=\frac{\la Z(y_k),Z(y_l)\ra}{(y_{k}y_{l})} \;\;\mbox{for}\;k\neq l, \qquad \HH^{\vee}_{kk}=-\frac{\la Z(y_{k}),\partial Z(y_{k})\ra}{\D y_{k}}.
\ee
It has been shown that this contribution is equivalent to a resultant of (roughly speaking) half the components of the map $Z^{I}$ to twistor space \cite{Cachazo:2013zc}.

All that remains is for us to deal with the correlator
\begin{equation*}
\left\la\prod_{i=1}^{n}h_{i} \prod_{j=1}^{n-d-2} Y_{j\;I}\rho^{I}[Y_{j},\tilde{\rho}]\right\ra.
\end{equation*}
The $\tilde{\rho}\rho$-system is fermionic and can only produce loops among the insertions, while the latter is bosonic and can produce trees (rooted at the $n$ wavefunction insertions) as well as loops.  This results in an overall counting of trees (since the loop diagrams cancel between boson and fermion systems) weighted by propagators, which is operationalized by the matrix-tree theorem (c.f., \cite{Feng:2012sy}). The role of this theorem was first explored in the context of twistor-strings in \cite{Adamo:2012xe}.  

The propagator for the $\tilde{\rho}\rho$-system is given by \eqref{rprop}, while the $YZ$-system has 
\be{YZprop}
\left\la Y_{I}(x)\;Z^{J}(\sigma)\right\ra=\delta^{J}_{I}\frac{\D x}{(x\sigma)}\prod_{r=1}^{d+1}\frac{(a_{r}\sigma)}{(a_{r}x)},
\ee
where the $d+1$ points $a_r\in\P^1$ account for the ambiguity of inverting the $\dbar$-operator on forms of weight $d$.  Combining the contributions of these systems via the matrix-tree theorem results in
\begin{equation*}
\left\la\prod_{i=1}^{n}h_{i} \prod_{j=1}^{n-d-2} Y_{j\;I}\rho^{I}[Y_{j},\tilde{\rho}]\right\ra=\left|\mathsf{H}\right|\prod_{i=1}^{n}h_i,
\end{equation*}
where $\mathsf{H}$ is a $(n-d-2)\times(n-d-2)$-matrix with off-diagonal entries
\be{HM1*}
\mathsf{H}_{jk}=\sum_{i=1}^{n}\sum_{l\neq i}\frac{\D x_{j}^{3/2}\;\D x_{k}^{3/2}}{(x_{j}x_{k})(x_{j}\sigma_{i})(x_{k}\sigma_{l})}\prod_{r=1}^{d+1}\frac{(a_{r}\sigma_{i})(a_{r}\sigma_{l})}{(a_{r}x_{j})(a_{r}x_{k})}\left[\frac{\partial}{\partial Z(\sigma_{i})},\frac{\partial}{\partial Z(\sigma_{l})}\right],
\ee
and diagonal entries
\be{HM2*}
\mathsf{H}_{jj}=\sum_{i=1}^{n}\frac{\D x_j^{3}}{(x_{j}\sigma_{i})^2}\prod_{r=1}^{d+1}\frac{(a_{r}\sigma_{i})^2}{(a_{r}x_{j})^2}\sum_{l\neq i}\frac{1}{(\sigma_{i}\sigma_{l})}\prod_{s=1}^{d+1}\frac{(a_{s}\sigma_{l})}{(a_{s}\sigma_{i})}\left[\frac{\partial}{\partial Z(\sigma_{i})},\frac{\partial}{\partial Z(\sigma_{l})}\right].
\ee
In computing these entries, care must be taken to properly account for potential short-distance singularities arising from same-site contractions in the $\rho\tilde{\rho}$-system.  These are compensated for by the anti-symmetry of the infinity twistor in a fashion similar to the calculation from \cite{Skinner:2013xp} which leads to the diagonal entries of \eqref{DHM*}.

Pulling all the pieces together, we at last obtain an expression for the genus zero scattering amplitudes of the twistor-string at degree $d$:
\be{SGZ1}
\cM_{n,d}=\int\frac{\prod_{a=0}^{d}\d^{4|8}U_{a}}{\mathrm{vol}\;\GL(2,\C)}\frac{|\mathsf{H}|}{|\mathsf{N}|^2}\frac{\left|\HH^{\vee}\right|}{|y_{1}\cdots y_{d}|^2}\prod_{i=1}^{n}h_i,
\ee
where the $\{U_{a}\}$ are the parameters of the holomorphic map $Z:\P^{1}\rightarrow\PT$.   

\medskip

Now, our goal is to show that $\cM_{n,d}$ given by \eqref{SGZ1} is independent of the locations of the PCOs, at $\{x_j\}$ and $\{y_k\}$.  In the latter case, the only $\{y_k\}$ dependence is in the ratio
\begin{equation*}
 \frac{\left|\HH^{\vee}\right|}{|y_{1}\cdots y_{d}|^2},
\end{equation*}
and it can be shown (c.f., \cite{Skinner:2013xp, Cachazo:2013zc}) that this has no poles in any of the $y_{k}$.  By Liouville's theorem, it follows that we can freely choose these points, so we can set $y_{k}=\sigma_{i}$.  Next, we must consider the dependence of $\cM_{n,d}$ on the PCO insertion points $x_j$. 

The strategy is once again to consider the potential singularities of $\cM_{n,d}$, but now with respect to the $x_{j}$.  These variables appear in the matrices $\mathsf{H}$ and $\mathsf{N}$, and inspection of \eqref{Nmat*}, \eqref{HM1*}, \eqref{HM2*} shows that potential singularities can develop whenever $x_j$ approaches one of the reference points $a_{r}$ or one of the vertex operator insertion points $\sigma_{i}$.  In the first case, these potential poles appear only in the entries of the matrix $\mathsf{H}$: diagonal entries have potential poles of order three, while off-diagonal entries have potential simple poles.

Without loss of generality, consider these potential singularities when the reference point $a_{0}\in\P^1$ coincides with $x_{j}$, the insertion point of the PCO $\widetilde{\Upsilon}_j$.   In the diagonal entries of $\mathsf{H}$, we are interested in the residue
\begin{equation*}
\mathrm{Res}_{a_{0}=x_j}\mathsf{H}_{jj}=\frac{1}{2}\lim_{a_{0}\rightarrow x_j}\frac{\d^{2}}{\d a_0^2}\left((a_{0}x_j)^3\;\mathsf{H}_{jj}\right).
\end{equation*}
To perform this calculation, it is convenient to assume that in \eqref{SGZ1} we have used standard momentum eigenstates of the form
\be{momeig}
h(Z(\sigma_i))=\int_{\C}\frac{\d t_{i}}{t^{3}_i}\bar{\delta}^{2}\left(\lambda_{i}^{\alpha}-t_{i}\lambda^{\alpha}(\sigma_i)\right)\e^{t_{i}[[\mu(\sigma_i)\tilde{\lambda}_i]]}.      
\ee
Using the definition of the diagonal entries \eqref{HM2}, this leads to
\be{resc1}
\mathrm{Res}_{a_{0}=x_j}\mathsf{H}_{jj}=4\sum_{i=1}^{n}\frac{\D x_{j}^{3}}{(x_{j}\sigma_{i})^2}(a_{0}\sigma_{i})\prod_{r=2}^{d+1}\frac{(a_{r}\sigma_{i})^2}{(a_{r}x_{j})^2}\sum_{l=1}^{n}[i\;l]\prod_{r=2}^{d+1}\frac{(a_{r}\sigma_{l})}{(a_{r}x_{j})},
\ee
which vanishes on the support of the wavefunctions \eqref{momeig} after integrating out the $\mu(\sigma)$-moduli in \eqref{PMSA}.  Here, we have been able to extend the second summation in $\mathsf{H}_{jj}$ over all vertex operators by extracting the residue.  

Likewise, the residue corresponding to the potential simple pole in the off-diagonal entries vanishes:
\be{resc2}
\mathrm{Res}_{a_{0}=x_j}\mathsf{H}_{jk}=\sum_{i,l=1}^{n}\frac{\D x_{j}^{3/2}\D x_{k}^{3/2}}{(x_{j}x_{k})}[i\;l]\prod_{r=2}^{d+1}\frac{(a_{r}\sigma_{i})(a_{r}\sigma_{l})}{(a_{r}x_{j})(a_{r}x_{k})}=0.
\ee
So in reality there are no singularities associated with the coincidence of the reference points $a_{r}$ and the PCO insertions.  All that remains is to check the potential singularities which can occur whenever the $x_j$ approach vertex operator insertion points $\sigma_i$.

Without loss of generality, we can look at the limit where $x_j$ approaches the operator insertion point $\sigma_{j+d+2}$.  The dependence on this limit is isolated in the ratio
\begin{equation*}
\frac{|\mathsf{H}|}{|\mathsf{N}|^2}
\end{equation*}
of \eqref{SGZ1}.  Expand the Slater determinant in the denominator as
\begin{multline*}
|\mathsf{N}|=S(\sigma_{n},x_{n-d-2})\left|
\begin{array}{cccccc}
\mathcal{Y}_{1}(\sigma_{1}) & \cdots & \mathcal{Y}_{d+2}(\sigma_{1}) & S(\sigma_{1},x_{1}) & \cdots & S(\sigma_{1},x_{n-d-3}) \\
\vdots & & \vdots & \vdots & & \vdots \\
\mathcal{Y}_{1}(\sigma_{n-1}) & \cdots & \mathcal{Y}_{d+2}(\sigma_{n-1}) & S(\sigma_{n-1},x_{1})& \cdots & S(\sigma_{n-1},x_{n-d-3})
\end{array}\right| + \cdots \\
=S(\sigma_{n},x_{n-d-2})S(\sigma_{n-1},x_{n-d-3})\cdots S(\sigma_{d+3},x_{1})\;|\sigma_{1}\cdots \sigma_{d+2}|+\cdots,
\end{multline*}
where the remaining terms contribute nothing to the ratio in the limit under consideration.  This means that we need to compute
\be{cslim1}
\lim_{x_{j}\rightarrow\sigma_{j+d+2}}\frac{|\mathsf{H}|}{|\sigma_{1}\cdots\sigma_{d+2}|^2}\prod_{j=1}^{n-d-2}S(\sigma_{j+d+2},x_{j})^{-2}.
\ee

Using the basic properties of determinants, we can absorb the product of propagators from the $\beta\gamma$-system into $|\mathsf{H}|$ to obtain
\be{cslim2}
\lim_{x_{j}\rightarrow\sigma_{j+d+2}}\frac{|\widehat{\mathsf{H}}|}{|\sigma_{1}\cdots\sigma_{d+2}|^2},
\ee
where the re-scaled matrix $\widehat{\mathsf{H}}$ has entries
\begin{equation*}
\widehat{\mathsf{H}}_{jk}\equiv \frac{\mathsf{H}_{jk}}{S(\sigma_{j+d+2},x_{j})S(\sigma_{k+d+2},x_{k})}.
\end{equation*}
From \eqref{HM1*}, \eqref{HM2*}, and \eqref{bgprop} we can see that the entries of $\widehat{\mathsf{H}}$ are actually just the entries of the familiar Hodges matrix $\HH$:
\be{cslim3}
\lim_{x_{j}\rightarrow\sigma_{j+d+2}}\widehat{\mathsf{H}}_{jk}=\frac{\sqrt{\D\sigma_{j+d+2}}\sqrt{\D\sigma_{k+d+2}}}{(\sigma_{j+d+2}\sigma_{k+d+2})}\left[\frac{\partial}{\partial Z(\sigma_{j+d+2})},\frac{\partial}{\partial Z(\sigma_{k+d+2})}\right],
\ee
\be{cslim4}
\lim_{x_{j}\rightarrow\sigma_{j+d+2}}\widehat{\mathsf{H}}_{jj}=-\sum_{k\neq j}\frac{\D\sigma_{j+d+2}}{(\sigma_{j+d+2}\sigma_{k+d+2})}\prod_{r=1}^{d+2}\frac{(a_{r}\sigma_{k+d+2})}{(a_{r}\sigma_{j+d+2})}\left[\frac{\partial}{\partial Z(\sigma_{j+d+2})},\frac{\partial}{\partial Z(\sigma_{k+d+2})}\right].
\ee
This means that the limit $x_{j}\rightarrow\sigma_{j+d+2}$ is smooth, so there are \emph{no} singularities in the position of the PCOs $\widetilde{\Upsilon}_j$.  By Liouville's theorem, this means that the $x_{j}$ can be freely chosen, and in particular we can set $x_{j}=\sigma_{j+d+2}$.

Hence, by setting the PCO insertion points to coincide with vertex operator insertions, we recover the familiar form of the Cachazo-Skinner formula for $\cM_{n,d}$ given by \eqref{CSi}.


\section{The Role of the Degeneration Parameter}
\label{QProps}

In this paper, we have taken for granted several claims about the role of the modulus $q$, which controls the worldsheet degeneration in the factorization limit.  These have included the relationship between $q$ and the momentum flowing through the factorization cut, and the fact that Wick contractions between external vertex operators on different factors are suppressed by $q$ as the degeneration is approached.  Although these claims are proven elsewhere in the literature, we use this appendix to review the salient points.  

We will use a slightly different model from \eqref{WF1} in order to describe the worldsheet near the separating boundary divisor.  This is mainly for convenience, since we need to keep track of explicit powers of the degeneration parameter.  We follow \cite{Cachazo:2012pz}, and model the worldsheet on a conic in $\P^{2}$:
\be{Qmod}
\Sigma=\left\{[x:y:z]|\;xy=s^{2}z^2\right\}.
\ee
This model for the rational worldsheet has the advantage of being global (in the sense that any rational curve can be expressed in this form) and manifestly projective.  The degeneration parameter $q$ used in the text is related to $s$ by $q=s^{2}$, so as $s\rightarrow 0$, the worldsheet $\Sigma$ degenerates into two factors $\Sigma_{L}$ and $\Sigma_{R}$ which intersect in a single point.  The model \eqref{Qmod} is therefore a projective version of \eqref{WF1} with the points $a\in\Sigma_L$ and $b\in\Sigma_R$ set to be the origin on each factor.

As usual, there is a natural homogeneous coordinate system on $\Sigma\cong\P^1$ given by $\sigma^{\alpha}=(\sigma^{0},\sigma^{1})$.  These are related to the projective coordinates of the embedding space $\P^{2}$ by
\begin{equation*}
 [x,y,z]=\left((\sigma^0)^2, (\sigma^{1})^{2},\frac{\sigma^{0}\sigma^{1}}{s}\right).
\end{equation*}
One advantage of the projective embedding \eqref{Qmod} is that it provides us with a natural set of homogeneous coordinates on the two factors $\Sigma_{L}$, $\Sigma_{R}$ in the degenerate limit:
\be{hcoords}
\sigma_{L}^{\alpha}=\sigma^{0}\left(\frac{\sigma^1}{s},\sigma^{0}\right), \qquad \sigma_{R}^{\alpha}=\sigma^{1}\left(\frac{\sigma^0}{s},\sigma^1\right).
\ee
Now, on the non-degenerate worldsheet $\Sigma$, we can define the affine coordinate $z=\sigma^{1}/\sigma^{0}$.  This in turn induces a choice of affine coordinates on the factors $\Sigma_L$, $\Sigma_R$:
\be{affine}
z_{L}=\frac{s}{z}, \qquad z_{R}=s\;z.
\ee

First, we prove that Wick contractions between external states on different factors are suppressed in the degenerate limit.  Consider a $n$-point worldsheet correlation function in twistor-string theory.  This can be in either the Berkovits-Witten or Skinner models, with external states represented by some BRST-closed vertex operators $\cV_i$; the details of these operators are unimportant.  Wick contractions between two of these vertex operators will take the general form
\begin{equation*}
 \cV(\sigma_i)\;\cV(\sigma_j)\sim \frac{\D\sigma_i\;\D\sigma_j}{(\sigma_i\sigma_j)}=\frac{\d z_i\;\d z_j}{z_i-z_j},
\end{equation*}
up to some proportionality factors accounting for the details of the Wick contraction.

For $|s|$ very small, suppose that $\cV_i$ is on $\Sigma_L$ and $\cV_j$ is on $\Sigma_R$.  Working with the affine coordinates \eqref{affine}, we see that
\begin{equation*}
\frac{\d z_i\;\d z_j}{z_i-z_j}=\frac{s\;z_{j\;R}}{z_{i\;L}^2}\frac{\d z_{i\;L}\;\d z_{j\;R}}{(s^2-z_{i\;L}z_{j\;R})},
\end{equation*}
which vanishes as $s\rightarrow 0$.  Hence, all such Wick contractions are suppressed as we approach the separating boundary divisor, as claimed in the text.  A similar calculation shows that Wick contractions between vertex operators on the same factor are non-vanishing as the boundary divisor is approached.

We conclude by demonstrating that the degeneration parameter is directly related to the momentum flowing through the factorization channel when momentum eigenstates are inserted.  For the Berkovits-Witten twistor-string, these momentum eigenstates take the form:
\be{meig}
a(Z(z_i);\lambda_i\tilde{\lambda}_i)=\int \frac{\d t_i}{t_i}\bar{\delta}^{2}\left(\lambda_i-t_i\lambda(z_i)\right)\;\e^{t_i[[\mu(z_i)\tilde{\lambda}_i]]},
\ee
while for the Skinner theory they are given by \eqref{momeig}.  In both cases, the external states have on-shell momentum $p_{i}^{\alpha\dot{\alpha}}=\lambda_{i}^{\alpha}\tilde{\lambda}_{i}^{\dot{\alpha}}$, and we represent the components of the map to twistor space in the usual way: $Z^{I}=(\lambda_{\alpha},\mu^{\dot{\alpha}},\chi^{a})$.  

Now, for every external state on $\Sigma_L$, the zero modes of the map $Z$ can be written in a re-scaled form \cite{Vergu:2006np, Cachazo:2012pz}:
\be{Qmap}
Z(z_L)=z^{d_L}\left(U_{\bullet}+\sum_{a=1}^{d_L}U_{a}z^{a}_{L}+\sum_{b=1}^{d_R}W_{b}\frac{s^{2b}}{z_{L}^{b}}\right),
\ee
where $U_{\bullet}$ gets mapped to the node in the degenerate limit and we have re-scaled
\begin{equation*}
U_{d_{L}-a}\rightarrow s^{-a}U_{a}, \qquad U_{d_{L}+b}\rightarrow s^{-b}W_{b},\qquad U_{d_L}\rightarrow U_{\bullet}.
\end{equation*}
In conjunction with the delta functions in the momentum eigenstates, \eqref{Qmap} tell us that upon performing the moduli integrals associated with the $\mu$-components of the map, we have
\begin{equation*}
P_{L}^{\alpha\dot{\alpha}}=\sum_{i\in L}\lambda_{i}^{\alpha}\tilde{\lambda}_{i}^{\dot{\alpha}}=U_{\bullet}^{\alpha}\sum_{i\in L}t_{i}\tilde{\lambda}_{i}^{\dot{\alpha}}+s^{2}W_{1}^{\alpha}\sum_{i\in L}t_{i}\frac{\tilde{\lambda}_{i}^{\dot{\alpha}}}{z_{i\;L}}+O(s^4), 
\end{equation*}
where $U_{\bullet}^{\alpha}$ is the $\lambda$-component of $U_{\bullet}$ and $W_{1}^{\alpha}$ is the $\lambda$-component of the map associated to the portion mapped to $\Sigma_R$.

This immediately tells us that
\be{Qmomentum}
P_{L}^{2}=s^{2}\;\left\la U_{\bullet} W_{1}\right\ra\;\left[\sum_{i\in L} t_{i}\tilde{\lambda}_{i}\:\sum_{j\in L} t_{j}\frac{\tilde{\lambda}_{j}}{z_{j\;L}}\right] +O(s^4).
\ee
Hence, as $s^2\rightarrow0$, the momentum flowing from the external states on $\Sigma_L$ becomes null, as expected for multiparticle factorization.  Furthermore, we see that one power of $P_{L}^{2}$ corresponds to one power of $s^{2}$; this is a consequence of the insertion of momentum eigenstates for the external data.  Translating this into $q= s^2$ used in the text, this proves the claim that in twistor-string theory there is a direct relationship between powers of $q$ and powers of $p^2$.  So a simple pole in $q$ corresponds to a propagator $p^{-2}$ going on-shell in the factorization limit, while a double pole corresponds to a propagator $p^{-4}$, and so forth.


\section{Multi-Trace Factorization}
\label{DTF}

In this appendix, we demonstrate that multi-trace terms in the gauge theory amplitudes of the Berkovits-Witten twistor-string have factorization channels corresponding to fourth-order propagators on momentum space.  In contrast to the argument presented in the main text, we perform this analysis at the level of the scattering amplitude itself, along the lines of \cite{Skinner:2010cz, Cachazo:2012pz}.  

Without loss of generality, we will consider a \emph{double-trace} contribution to a $n$-point, degree $d$ amplitude in the twistor-string.  At the level of the worldsheet correlation function, this is given by:
\be{DTF1}
\tr\left(\mathsf{T}^{a_1}\cdots\mathsf{T}^{a_{n_L}}\right)\;\tr\left(\mathsf{T}^{b_1}\cdots\mathsf{T}^{b_{n_R}}\right)\int \frac{\prod_{a=0}^{d}\d^{4|4}U_{a}}{\mathrm{vol}\;\C^{*}}\left\la\prod_{\alpha=1}^{n-3}(b_{\alpha}|\mu_{\alpha})\;\prod_{i=1}^{n_L}\widetilde{\cV}_{a_i}\prod_{j=1}^{n_R}\widetilde{\cV}_{b_j}\right\ra.
\ee
We write the double trace explicitly to indicate that in the worldsheet current algebra, we exclude Wick contractions between the set of vertex operators $\{\cV_i^{a}\}_{i=1,\ldots,n_L}$ and $\{\cV^{a}_{j}\}_{j=1,\ldots,n_R}$.  This correlator can be evaluated without difficulty to give:
\be{DTF2}
\tr\left(\mathsf{T}^{a_1}\cdots\mathsf{T}^{a_{n_L}}\right)\;\tr\left(\mathsf{T}^{b_1}\cdots\mathsf{T}^{b_{n_R}}\right)\int \frac{\prod_{a=0}^{d}\d^{4|4}U_{a}}{\mathrm{vol}\;\GL(2,\C)}\prod_{i=1}^{n_L}\frac{a(Z_i)\wedge\D\sigma_i}{(\sigma_i\sigma_{i+1})}\prod_{j=1}^{n_R}\frac{a(Z_j)\wedge\D\sigma_j}{(\sigma_j\sigma_{j+1})},
\ee
where each product is assumed to be defined modulo $n_L$ or $n_R$ respectively.

We focus on the degeneration of the worldsheet which preserves the double-trace structure; in particular, assume that in the degenerate limit the vertex operators $\{\widetilde{\cV}_{a_i}\}_{i=1,\ldots,n_L}$ are located on $\Sigma_L$ and $\{\widetilde{\cV}_{b_j}\}_{j=1,\ldots,n_R}$ are located on $\Sigma_R$, as our notation suggests. As in Appendix \ref{QProps}, we use \eqref{Qmod} for our model of the worldsheet.  We want to isolate the dependence of \eqref{DTF2} on the degeneration parameter $s$ as we approach the factorization limit.

To do this, we must translate \eqref{DTF2} into an expression in terms of affine coordinates adopted to each factor; these are given naturally by \eqref{affine}.  First, consider the measure on the vertex operator locations
\begin{equation*}
 \frac{\prod_{i=1}^{n}\D\sigma_i}{\mathrm{vol}\;\SL(2,\C)}=\frac{\prod_{i=1}^{n}\d z_i}{\mathrm{vol}\;\SL(2,\C)}.
\end{equation*}
Translating this into the local affine coordinates gives a measure near the boundary divisor \cite{Cachazo:2012pz}:
\be{posmeasure}
\frac{s^{n_{L}-n_{R}-4}\d s^{2}}{\prod_{i=1}^{n_L}z_{i\;L}^{2}}\left(\frac{\prod_{i=1}^{n_{L}+1}\d z_{i\;L}}{\mathrm{vol}\;\SL(2,\C)}\right)\left(\frac{\prod_{j=1}^{n_{R}+1}\d z_{j\;R}}{\mathrm{vol}\;\SL(2,\C)}\right).
\ee
Here, $\d z_{n_{L}+1\;L}=\d z_{a\;L}$ and $\d z_{n_{R}+1\;R}=\d z_{b\;R}$ are the measures associated to the new marked points $a\in\Sigma_L$ and $b\in\Sigma_R$.  We also need to translate the denominator factors in \eqref{DTF2} which arose from the worldsheet current algebra:
\be{cameasure1}
\prod_{i=1}^{n_L}\frac{1}{z_i-z_{i+1}}=\frac{1}{s^{n_L}}\prod_{i=1}^{n_L}\frac{z_{i\;L}z_{i+1\;L}}{z_{i\;L}-z_{i+1\;L}},
\ee
\be{cameasure2}
\prod_{i=1}^{n_L}\frac{1}{z_i-z_{i+1}}=s^{n_R}\prod_{i=1}^{n_L}\frac{1}{z_{i\;R}-z_{i+1\;R}}.
\ee

The only other place where the degeneration parameter can enter in \eqref{DTF2} is through the measure on the map moduli or the twistor wavefunctions $a(Z_i)$.  However, since the twistor space $\PT\subset\P^{3|4}$ is Calabi-Yau the former is invariant under the rescalings used to define the affine coordinates $z_L, z_R$.  Furthermore, the wavefunctions themselves are homogeneous: $a\in H^{0,1}(\PT,\cO\otimes\mathfrak{g})$, so there is no new $s$-dependence introduced through the external states.

This means that as the factorization limit is approached, the dependence of \eqref{DTF2} on the moduli looks like
\be{DTF3}
\frac{\d s^{2}}{s^4}+O(s^{-2}),
\ee
while the worldsheet measure \eqref{posmeasure} and map measure factorize appropriately.  As discussed in Appendix \ref{QProps}, when momentum eigenstates are inserted for the external wavefunctions in \eqref{DTF2}, this has the structure of a $p^{-4}$ pole in momentum space.  In the language used in the text, \eqref{DTF3} is a $q^{-2}$ double pole as we approach the boundary divisor.  This confirms that the sub-leading trace contributions to the Berkovits-Witten twistor-string factorize as expected for conformal supergravity.

\bibliography{tsfactor}
\bibliographystyle{JHEP}

\end{document}